\documentclass[sigconf]{acmart} 
\AtBeginDocument{%
  \providecommand\BibTeX{{%
    \normalfont B\kern-0.5em{\scshape i\kern-0.25em b}\kern-0.8em\TeX}}}

\setcopyright{acmcopyright}
\copyrightyear{2024}
\acmYear{2024}
\setcopyright{acmlicensed}\acmConference[WSDM '24]{Proceedings of the 17th ACM International Conference on Web Search and Data Mining}{March 4--8, 2024}{Merida, Mexico}
\acmBooktitle{Proceedings of the 17th ACM International Conference on Web Search and Data Mining (WSDM '24), March 4--8, 2024, Merida, Mexico}
\acmPrice{15.00}
\acmDOI{10.1145/3616855.3635826}
\acmISBN{979-8-4007-0371-3/24/03}

\usepackage{soul}
\usepackage{url}
\usepackage[switch]{lineno}
\usepackage[utf8]{inputenc}
\usepackage{graphicx}
\usepackage{amsmath}
\usepackage{amsthm}
\usepackage{algorithm}
\usepackage{algorithmicx}
\usepackage{algpseudocode}
\usepackage{enumitem}
\usepackage{moresize}
\allowdisplaybreaks

\usepackage{epsfig}
\usepackage{amsfonts}
\usepackage{multirow}
\usepackage{multicol}
\usepackage{subfigure}
\usepackage{comment}
\usepackage{color}
\usepackage[utf8]{inputenc}

\usepackage{balance}

\usepackage{thm-restate,thmtools}

\newtheorem{definition}{Definition}

\makeatletter
\algrenewcommand\ALG@beginalgorithmic{\small}
\makeatother

\begin{document}

%\title{Influence-based Evasion Attacks to Graph Neural Networks}
%\title{Evasion Attacks to Graph Neural Networks via Influence Function}
\title{Efficient, Direct, and Restricted Black-Box Graph Evasion Attacks to Any-Layer Graph Neural Networks via Influence Function}
\graphicspath{ {./figure/} }
% \author{Binghui Wang$^{*}$,  Minhua Lin$^{*}$, Tianxiang Zhou$^{*}$, Pan Zhou, Ang Li, Meng Pang, Hai Li, Yiran Chen \\
% %HUST\\
% }

\author{Binghui Wang}
\email{bwang70@iit.edu}
\affiliation{
\institution{Illinois Institute of Technology}
\city{Chicago}
\country{USA}}
\authornote{These authors contribute equally to this paper}

\author{Minhua Lin}
\email{mfl5681@psu.edu}
\affiliation{
\institution{The Pennsylvania State University}
\city{State College}
\country{USA}}
\authornotemark[1]

\author{Tianxiang Zhou}
\email{tianxiangzhou@hust.edu.cn}
\affiliation{
\institution{Huazhong University of Science and Technology}
\city{Wuhan}
\country{China}}
\authornotemark[1]

\author{Pan Zhou}
\email{panzhou@hust.edu.cn}
\affiliation{
\institution{Huazhong University of Science and Technology}
\city{Wuhan}
\country{China}}

\author{Ang Li}
\email{angliece@umd.edu}
\affiliation{
\institution{University of Maryland}
\city{College Park}
\country{USA}}

\author{Meng Pang}
\email{mengpang@ncu.edu.cn}
\affiliation{
\institution{Nanchang University}
\city{Nanchang}
\country{China}}

\author{Hai Li}
\email{hai.li@duke.edu}
\affiliation{
\institution{Duke University}
\city{Durham}
\country{USA}}

\author{Yiran Chen}
\email{yiran.chen@duke.edu}
\affiliation{
\institution{Duke University}
\city{Durham}
\country{USA}}

%\begin{document}
\begin{abstract}

Graph neural network (GNN), the mainstream method to learn on graph data, is vulnerable to graph evasion attacks, where an attacker slightly perturbing the graph structure can fool trained GNN models. Existing work has at least 
one of the following drawbacks: 
1) limited to directly attack two-layer GNNs; 
2) inefficient; and 3) impractical, as they need to know 
full or part of GNN model parameters. 

We address the above drawbacks and propose an influence-based \emph{efficient, direct, and restricted black-box} evasion attack to \emph{any-layer} GNNs.
Specifically, we first introduce two influence functions, i.e., feature-label influence and label influence, that are defined on GNNs and label propagation (LP), respectively. Then we observe that GNNs and LP
are strongly connected in terms of our defined influences.  
Based on this, we can then reformulate the evasion attack to GNNs as calculating label influence on LP, which is \emph{inherently} applicable to any-layer GNNs, while no need to know information about the internal GNN model.
Finally, we propose an efficient algorithm to calculate label influence. Experimental results on various graph datasets show that, compared to state-of-the-art white-box attacks, our attack can achieve comparable attack performance, but has a 5-50x speedup when attacking two-layer GNNs. Moreover, our attack is effective to attack multi-layer GNNs\footnote{Source code and full version is in the link: \url{https://github.com/ventr1c/InfAttack}}.
\end{abstract}


%Keywords. The author(s) should pick words that accurately describe
% the work being presented. Separate the keywords with commas.

\settopmatter{printfolios=true}

\maketitle
\section{Introduction}

Learning with graph data, such as social networks, biological networks, financial networks, has drawn continuous attention recently. Graph neural network (GNN) has become the  mainstream methodology for representation learning on graphs.  
GNN was first introduced in~\cite{scarselli2008graph}, which extended conventional neural network to process graph data. 
Then, various GNN methods have been proposed and achieved state-of-the-art performance in many graph-related tasks such as node classification~\cite{kipf2017semi,velivckovic2018graph,xu2019powerful}, graph classification~\cite{hamilton2017inductive,gilmer2017neural}, and link prediction~\cite{zhang2018link}. 
However, recent works~\cite{dai2018adversarial,zugner2018adversarial,wu2019adversarial,zugner2019adversarial,wang2019attacking,xu2019topology,sun2020adversarial,ma2020towards,fan2021reinforcement,mu2021hard,wang2022bandits,wang2023turning} show that GNNs are vulnerable to graph evasion attacks---Given a target node and a trained GNN model, an attacker slightly perturbing the graph structure\footnote{An attacker can also perturb node features to perform the attack. However, structure perturbation is shown to be much more effective than node feature perturbation~\cite{zugner2018adversarial}.} (e.g., add new edges to or delete existing edges from the graph)
can make the GNN model misclassify the target node. 
Existing attacks to GNNs can be roughly classified as 
\emph{optimization-based} attacks~\cite{zugner2018adversarial,xu2019topology,wu2019adversarial,wang2022bandits} and \emph{reinforcement learning (RL)}-based attacks~\cite{dai2018adversarial,sun2020adversarial,fan2021reinforcement}. 

In this paper, we focus on optimization-based attacks, 
as they are shown to be more effective~\cite{zugner2018adversarial}.
Optimization-based attacks first formulate the graph evasion attack as a binary optimization problem, which is challenging to solve, and then design approximate algorithms to solve a tractable optimization problem. 
Although achieving promising attack performance, existing optimization-based attacks have one or more of the below key limitations:
\begin{itemize}[leftmargin=*]

\item First, most of the existing attacks 
% they 
need to know the full/partial GNN model parameters, which is  unrealistic in many real-world applications, e.g., when GNN models are confidential due to their commercial value and are deployed as an API. 
Thus, the practicability of the existing attacks are limited.  
Further, they are mainly designed to attack \emph{two-layer GNNs}, while GNNs are multi-layer in essence. 
To attack multi-layer GNNs, they often first \emph{indirectly} attack a surrogate two-layer GNN model, and then transfer the attack to the target multi-layer GNN. However, this strategy is 
% introduce surrogate models that are 
not effective enough (See Figure~\ref{fig:GCN_4layer_attack} in Section~\ref{sec:eval}). 

\item Second, they are not efficient, as they involve intensive computation, i.e., by multiplying GNN model parameters of different layers and with node feature matrix. 
If a GNN has many layers,
such computation can be a bottleneck, especially for attackers who have limited computational resources or/and want to perform real-time attacks. For example, 
many fraud detection systems, such as detecting fake users in social networks and detecting anomalies from system logs, are updated frequently in order to reduce the loss caused by the evasion attacks’ malicious activities. In these scenarios, efficiency is 
a major concern for the attack and an attacker performing efficient attacks is necessary, as otherwise the detection system may have already updated and identified the attack’s malicious patterns before the attack is implemented. 
\end{itemize}

\noindent {\bf Our work:} 
We aim to address the above limitations in this paper. 
To this end, we propose an optimization-based evasion attack against any-layer GNNs based on influence function~\cite{koh2017understanding}---a completely different perspective from the existing works. 
Our influence-based attack is motivated by the strong connection between GNNs and label propagation (LP)~\cite{zhu2003semi}. 
Specifically, we first introduce two influence functions, i.e., feature-label influence and label influence, that are defined on 
GNNs and LP, respectively.
Then, we prove that our label influence defined on LP is equivalent to feature-label influence on a particular well-known type of GNN, called Graph Convolutional Network (GCN)~\cite{kipf2017semi} (and its linearized version Simple Graph Convolutional (SGC)~\cite{wu2019simplifying}).   
Based on this connection, we reformulate the evasion attack against GNNs to be related to calculating label influence on LP. 
As our influences are designed for any-layer GNNs, 
our attack is inherently applicable to attack any-layer GNNs. 
Note that label influence can be computed easily and we also design an efficient algorithm to compute it. 
Further, as our influence-based attack does not need to know any information about the GNN model (except the target node's neighboring information), 
it is 
a more practical (restricted black-box) attack.  
Finally, we evaluate our attack against GCN/SGC on three benchmark graph datasets.
Compared to the state-of-the-art white-box attacks against two-layer GCN/SGC, our attack can achieve comparable attack performance but has a 5-50x speedup. 
Our attack is more effective to attack multi-layer GCN/SGC. 
For instance, our attack achieves a $93\%$ attack success rate, when perturbing 4 edges per target node on Cora, while the surrogate model based attack only has $80\%$ attack success rate. 
As a by product, our attack also shows promising transferability to attack other GNNs, and is more effective than existing black-box attacks.  

Our contributions can be summarized as follows: 
\begin{itemize}[leftmargin=*]
%\vspace{-1.5mm}
\item We propose graph evasion attacks to GNNs based on influence function, which is a completely new perspective. 
%\vspace{-1mm}
\item Our attack is effective, direct, efficient, and practical. 
%\vspace{-1mm}
\item Our attack has promising transferability. %to other GNNs.

\end{itemize}
%\vspace{-6mm}
\section{Related Work}
\label{sec:related}

\noindent {\bf Attacks to graph neural networks.}
Existing attacks to GNNs can be classified as graph \emph{poisoning attacks}~\cite{zugner2018adversarial,dai2018adversarial,zugner2019adversarial,xu2019topology,takahashi2019indirect,liu2019unified,sun2020adversarial,zhang2020backdoor,zhang2020adversarial,xi2021graph, dai2023unnoticeable} and \emph{evasion attacks}~\cite{dai2018adversarial,zugner2018adversarial,wu2019adversarial,ma2020towards}.  
In poisoning attacks, an attacker  modifies the graph structure during the training process such that the trained GNN model has a low prediction accuracy on testing nodes. 
For instance, 
\citet{xu2019topology} developed a topology poisoning attack based on gradient-based optimization. 
Evasion attacks can be classified as untargeted attacks and targeted attacks, where the latter is more challenging. 
Given a target node and a trained GNN model, targeted attack means an attacker aims to perturb the graph structure such that the GNN model misclassifies the target node to be a target label, while untargeted attack misclassifies the target node to be an arbitrary label different from the target node's label.    
For instance, \citet{dai2018adversarial} leveraged reinforcement learning techniques to design non-targeted evasion attacks to both graph classification and node classification.
\citet{zugner2018adversarial} proposed a targeted evasion attack, called Nettack, against two-layer GCN  
and achieved the state-of-the-art attack performance.  
Specifically, Nettack learns a surrogate linear model of GCN by removing the ReLU activation function and by defining a graph structure
preserving perturbation that constrains the difference between the node degree distributions of the graph before and after attack. 
Our label influence-based attack is a targeted evasion attack. 

Most of the existing GNN attacks are white/gray-box. Recently, two black-box attacks to GNNs~\cite{mu2021hard,wang2022bandits} have been proposed. For instance, Wang et al.~\cite{wang2022bandits} 
formulate the black-box attack to GNNs as an online optimization with bandit feedback. The original problem is NP-hard and they then propose 
an online attack based on (relaxed) bandit convex optimization which is proven to be {sublinear} to the query number.  Our attack is a restricted black-box attack, where the attacker only needs to know the target node's neighbors. 

\noindent {\bf Attacks to other graph-based methods.}
Besides attacking GNNs, other adversarial attacks against graph data include attacking graph-based clustering~\cite{chen2017practical}, graph-based collective classification~\cite{torkamani2013convex,wang2019attacking}, graph embedding~\cite{dai2018adversarialnet,sun2018data,chen2018fast,bojchevski2019adversarial,chang2020restricted}, community detection~\cite{li2020adversarial}, etc. 
For instance, \citet{chen2017practical} proposed a practical attack against spectral clustering, which is a well-known graph-based clustering method. 
~\citet{wang2019attacking} designed an optimization-based attack against the collective classification method, called linearized belief propagation, by modifying the graph structure. 

\noindent {\bf Defending against graph perturbation attacks.}
Existing defenses against the graph perturbation attacks can be classified as empirical defenses~\cite{wu2019adversarial,xu2019topology,zhu2019robust,tang2020transferring,entezari2020all,tao2021adversarial} and provable defenses~\cite{bojchevski2020efficient,jin2020certified,bojchevski2019certifiable,wang2021certified, lin2023certifiably}. The empirical defenses are shown to be easily  broken by stronger/adaptive  attacks~\cite{geisler2020reliable,mujkanovic2022defenses}. 
Provable defenses 
study certified robustness of GNNs against the worst-case graph perturbation attacks.  
For instance, Wang et al.~\cite{wang2021certified} design a randomized smoothing-based provable defenses that achieves a tight certified robustness, when there are no assumptions about the GNN model. \cite{wang2021certified} achieves the state-of-the-art provable defense performance.

% \vspace{-3mm}
\section{Background and Problem Definition}

{\bf Graph Neural Network.}
Let $G=(\mathcal{V}, \mathcal{E}, \mathbf{X})$ be a graph, where $u \in \mathcal{V}$ is a node, $(u, v) \in \mathcal{E}$ is an edge between $u$ and $v$, and $\mathbf{X} = [\mathbf{x}_1; \mathbf{x}_2; \cdots; \mathbf{x}_{n}] \in \mathbb{R}^{n \times d} $ is the node feature matrix. 
We denote $\mathbf{A} = [\mathbf{a}_1; \mathbf{a}_2; \cdots; \mathbf{a}_{n}] \in \{0,1\}^{n \times n}$ 
as the adjacency matrix, 
 $d_u$ and $\Gamma_u$ as $u$'s node degree and the neighborhood set of $u$ (including self-loop $(u,u)$). 
We consider GNNs for node classification in this paper. In this context, 
 each node $u$
 has a label $y_u$ 
from a label set $\mathcal{Y} = \{1, 2, \cdots, C \}$. 
Given a set of $\mathcal{V}_L \subset \mathcal{V}$ labeled nodes 
$\{(\mathbf{x}_u, y_u)\}_{u \in \mathcal{V}_L}$ 
as the training set, 
GNN for node classification is to take the graph $G$ and labeled nodes as input
and learn a node classifier that maps each node $u \in \mathcal{V} \setminus \mathcal{V}_L$ to a class $y \in \mathcal{Y}$. 
In this paper, we focus on Graph Convolutional Network (GCN)~\cite{kipf2017semi}, a widely used type of GNN, and its special case Simple Graph Convolution (SGC)~\cite{wu2019simplifying}.

\noindent \emph{GCN.} 
GCN is motivated by spectral graph convolution~\cite{defferrard2016convolutional}. 
Suppose GCN has $K$ layers. We denote node $v$'s representation in the $k$-th layer as $\mathbf{h}_v^{(k)}$, where $\mathbf{h}_v^{(0)} = \mathbf{x}_v$. 
Then, GCN has the following form to update the node representation: 
{
%\small
\begin{align}
    \label{eqn:gcn}
    \mathbf{h}_v^{(k)} = \textrm{ReLU}\Big( \mathbf{W}^{(k)}
    \big( \sum\nolimits_{u \in \Gamma_v} d_u^{-1/2} d_v^{-1/2} \mathbf{h}_{u}^{(k-1)} \big) \Big).
\end{align}
}%
A node $v$'s final representation $\mathbf{h}_v^{(K)} \in \mathbb{R}^{|\mathcal{Y}|}$ can capture the structural information of all nodes within $v$'s $K$-hop neighbors. 
Moreover, the final node representations of training nodes are used for training the node classifier.
Specifically, let $\Theta = \{\mathbf{W}^{(1)}, \mathbf{W}^{(2)}, \cdots, \mathbf{W}^{(K)} \} $ be the model parameters and $v$'s output be 
$ f_{\Theta} (\mathbf{A})_v = 
\textrm{softmax}(\mathbf{h}_v^{(K)}) \in \mathbb{R}^{|\mathcal{Y}|}$, where $f_{\Theta} (\mathbf{A})_{v,y}$ indicates the probability of node $v$ being class $y$. %(under $\Theta$). 
Then, $\Theta$ are learnt by minimizing the cross-entropy loss on the 
% outputs of the 
training nodes $\mathcal{V}_L$, i.e., $\Theta^* = \arg \min\nolimits_{\Theta} - \sum_{v \in \mathcal{V}_L} 
    \ln f_{\Theta} (\mathbf{A})_{v,y}.$ 
With the learnt $\Theta^*$, we can predict 
the label for each unlabeled nodes $u \in \mathcal{V} \setminus \mathcal{V}_L$ as $\hat{y}_u = \arg \max_{y} \, f_{\Theta^*}(\mathbf{A})_{u,y}$.

\noindent \emph{SGC.} 
SGC is a linearized version of GCN. 
Specifically, 
its node representation is updated as follows: 
\begin{align}
\label{eqn:sgc}
\small
\mathbf{h}_v^{(k)} = \mathbf{W}^{(k)}
    \big( \sum\nolimits_{u \in \Gamma_v} d_u^{-1/2} d_v^{-1/2} \mathbf{h}_{u}^{(k-1)} \big).
\end{align}
SGC has shown to have comparable node classification performance with GCN, but is much more efficient than GCN.

\noindent {\bf Label Propagation (LP).}
LP is a conventional semi-supervised node classification method without training.   
The key idea behind LP is that two nodes having a high similarity (e.g., connected nodes in a graph) are likely to have the same label. 
Thus, LP iteratively propagates labels among the graph to unlabeled nodes based on node-pair similarity. 
Let $\mathbf{y}_v \in \mathbb{R}^{|\mathcal{Y}|}$ be node $v$'s initial label vector (For notation reason, one should note that $y_v$ is $v$'s categorical label). 
For instance, $\mathbf{y}_v$ can be $v$'s one-hot label vector if $v$ is a labeled node, and $\mathbf{y}_v = \mathbf{0}$, otherwise. 
Then, LP is formulated as follows: 
{
\begin{equation}
    \label{eqn:LP}
    \mathbf{y}_v^{(k)} = \sum\nolimits_{u \in \Gamma_v} d_u^{-1/2} d_v^{-1/2} \mathbf{y}_{u}^{(k-1)}, \quad \mathbf{y}_v^{(0)} = \mathbf{y}_v.
\end{equation}
}%
With $K$ iterations, an unlabeled node $u$ is predicted to be class $c$, if $c = \arg\max_{i} {y}^{(K)}_{u, i}$. 

\noindent \emph{GNN vs. LP:} Viewing Eqn (\ref{eqn:LP}) and Eqns  (\ref{eqn:gcn}) and (\ref{eqn:sgc}), we observe that LP and GNNs  have similar iterative processes: LP propagates node labels $\mathbf{y}_v$, while GNNs propagate node features $\mathbf{x}_v$.  
The key difference is that LP does not involve model parameters, while GNN involves multiplying the parameter matrix $\mathbf{W}^{(k)}$ in each $k$-th layer.    

\noindent {\bf Problem Definition.} We consider targeted evasion attacks\footnote{As untargeted attacks are less powerful than targeted attacks, we only consider targeted attacks in this paper for simplicity.} to GNNs. 
Suppose we are given a trained GNN model $f_{\Theta^*}$ for node classification. 
We assume $v$ is the \emph{target node} and $c$ is the \emph{target label}. 
We consider an attacker can perturb the graph structure 
(i.e., add new edges to or delete existing edges from the graph) 
in order to make $f_{\Theta^*}$ misclassify the target node $v$ to be the target label $c$.
We call the modified edges by the attacker as \emph{attack edges}.
In particular, we consider a practical \emph{direct attack}~\cite{zugner2018adversarial}, where an attacker can only modify the edge status between $v$ and other nodes in the graph, while cannot modify the edge status among other nodes. 
We denote the perturbed graph as $\tilde{G}$ (with the perturbed adjacency matrix $\tilde{\mathbf{A}}$) after the attack and the attack budget as $\Delta$, i.e., at most $\Delta$ edges can be perturbed for the target node. 
Then, the objective function of targeted evasion attacks to GNNs is formally defined as:
{
\begin{align}
    \label{eqn:attack}
    & \max\nolimits_{\tilde{\mathbf{A}}_v} \big( f_{\Theta^*}(\tilde{\mathbf{A}})_{v, c}- f_{\Theta^*}(\tilde{\mathbf{A}})_{v,y_v} \big) \Leftrightarrow \max\nolimits_{\tilde{\mathbf{A}}_v} \big( [\tilde{\mathbf{h}}_v^{(K)}]_c - [\tilde{\mathbf{h}}_v^{(K)}]_{y_v} \big), \notag \\
    & \, s.t., \quad \sum\nolimits_{s} |\tilde{A}_{v,s} - A_{v,s}| \leq \Delta,
\end{align}
}%
where 
$\tilde{\mathbf{h}}_v^{(K)}$ is $v$'s representation on the perturbed graph $\tilde{G}$.

A target node is called a success to attack the GNN model if the value of the attack's objective function is larger than 0, under the attack budget. 
Note that 
Equation (\ref{eqn:attack}) is a binary optimization problem and is challenging to solve in practice. 
\citet{zugner2018adversarial} proposed an optimization-based attack method, called Nettack, against two-layer GCN.  
Specifically, Nettack attacked a substitute GNN model (actually SGC) that removed the ReLU activation function in GCN. 
Nettack has achieved state-of-the-art attack performance. However, it is inefficient as it involves dense matrix multiplication (i.e., model parameters multiply node features); it also needs to know model parameters $\Theta^*$, and can only attack two-layer GNNs.

%\vspace{-2mm}
%\section{Proposed Influence-based Evasion Attack}
\section{Influence-based Evasion Attack}

In this section, we propose our evasion attack against GNNs via influence function. 
In contrast to existing optimization-based attacks
that only focus on 
two-layer GNNs, our attack 
is applicable to any-layer GNNs. 
Specifically, we first define 
two influence functions associated with GNNs and LP, respectively, and build an equivalence relation between 
GNNs 
and LP with the defined influences. 
Next, we reformulate the attack objective function 
as relating to label influence defined on LP.
Finally, we design an efficient algorithm to calculate label influence and realize our attack.  

\subsection{Equivalence between GNNs and LP in terms of Influence}

\subsubsection{Motivation} Due to GNN’s complex network structure, existing optimization-based evasion attacks can only attack two-layer GNNs \emph{directly}. 
However, we note that LP has a similar iterative process to GNNs, but it has good properties, e.g., LP does not involve model parameters. 
Motivated by this, we aim
to discover an equivalence relation between LP and GNNs, such that the challenging problem of attacking multi-layer GNNs can be converted to a relatively easier problem by leveraging good properties of LP.
We notice that influence function~\cite{koh2017understanding,xu2018representation} is an appropriate tool to bridge the gap, and our purpose is to explore equivalent influence functions defined on LP and on GNNs, respectively. As the attacker's goal is to change the target node’s label, we thus need to define influences associated with the node label. As LP propagates node labels, we can naturally design the \emph{label influence} function (see Equation (\ref{label_influence})).
In addition, GNNs involve propagating node features. In order to also leverage node labels, we integrate both node features and labels and design the \emph{feature-label influence} function (see Equation (\ref{feature_influence})). 
Next, we introduce our influence functions.

%\vspace{-2mm}
\subsubsection{Influence function}
Given two nodes $u$ and $v$, an influence of $u$ on $v$ indicates how the output (e.g., final node representation in GNNs or estimated node label in LP) of $v$ changes if the input of $u$ is slightly perturbed. Inspired by~\cite{koh2017understanding,xu2018representation}, we 
define the following feature-label influence on GNN and label influence on LP. % respectively. 

% \vspace{-4mm}
\begin{definition}[Feature-label influence]
We define the feature-label influence of node $u$ on node $v$ associated with $u$'s label on a $K$-layer GNN as follow:
{
\small
\begin{equation}
\label{feature_influence}
    I_{fl}(v,u;K) 
    = \Big\| \Big[\frac{\partial \mathbf{h}_{v}^{(K)}}{\partial \mathbf{h}_{u}^{(0)}} \cdot \mathbf{h}_{u}^{(0)}  \Big]_{y_u} \Big\|_1
    = \mathbf{1}_{y_u}^T \cdot \frac{\partial \mathbf{h}_v^{(K)}}{\partial \mathbf{h}_u^{(0)}} \cdot \mathbf{h}_{u}^{(0)},
    %\mathbf{1}},
\end{equation}
}%
where $\mathbf{1}_{y_u}=[y_1,y_2,\cdots,y_n]$ is an indicator vector where $y_i=1$ if $i=u$ and $y_i=0$, otherwise; $\| \cdot \|_1$ is the vector $\ell_l$-norm; $T$ is a transpose; and
$\mathbf{h}_{u}^{(0)}=\mathbf{x}_u$ is $u$'s node features. %$\mathbf{x}_u$.
\end{definition} 

\begin{definition}[Label influence]
We define the label influence of node $u$ on node $v$ after $K$ iterations of label propagation 
as follows:
{
\small
\begin{align}
\label{label_influence}
    I_l(v,u;K) = \frac{\partial y_v^{(K)}}{\partial y_u^{(0)}}. 
    % \vspace{-2mm}
\end{align}
}%
\end{definition}

Then, we have the following theorem showing the equivalence between GNN and LP in terms of influence. 

\begin{theorem}
\label{thm:relate}
If the GNN is a GCN/SGC, then: 
\begin{equation}
\label{influence_relationship}
    I_{fl}(v,u;K) = C \cdot I_l(v,u;K),  
\end{equation}
where $C= \rho \mathbf{1}_{y_u}^T [{\prod_{l=1}^{K} \mathbf{W}^{(l)}}] \mathbf{x}_u$ is constant related to GNN model parameters $\Theta=\{\mathbf{W}^{(l)}\}$ and $u$'s  node features $\mathbf{x}_u$.  
\end{theorem}
Theorem~\ref{thm:relate} reveals that: given arbitrary node $v$, the feature-label influence defined on $K$-layer GCN/SGC of any other node $u$ on the node $v$ and the label influence defined on $K$-iteration LP of node $u$ on the node $v$ are equal (with a constant multiplier difference).

\subsection{Reformulate Evasion Attacks to Any-Layer GNNs as Calculating Label Influence on LP}
\label{sec:relation}
Based on our influence functions and Theorem~\ref{thm:relate}, we can first  
restate the challenging problem of attacking $K$-layer GNNs in Equation (\ref{eqn:attack}) in the form of feature-label influence, and further convert it to an equivalent problem related to label influence on LP. Before going into details, we first introduce 
the following lemma: 
\begin{restatable}[Xu et al.\cite{xu2018representation}]{lem}{gcninfluencelemma} 
\label{lemma:gcn_influence}
 Given a $K$-layer GCN. 
 Assume all paths in the computation graph of the GCN model are activated (i.e., via ReLU) with the same probability of success $\rho$. 
Then, 
{
\small
\begin{equation}
\label{derivate_h_v_k}
    \frac{\partial \mathbf{h}_v^{(K)}}{\partial \mathbf{h}_u^{(0)}} = \rho \sum_{p=1}^{\varPsi_{v \rightarrow u}}{\prod_{l=K}^1{a_{v_p^{l},v_p^{l-1}} \cdot \mathbf{W}^{(l)}}},
\end{equation}
}%
where  
$\varPsi_{v \rightarrow u}$ is the total number of paths $[v_p^K,v_p^{K-1},$ $\cdots,v_p^{1},v_p^{0}]$ of length $K+1$ from node $v$ to the node $u$ with $v_p^K=v$ and $v_p^0=u$. 
For $l=1,\cdots,K$, 
$v_p^{l-1}\in N\left( v_{p}^{l} \right) $, 
$a_{v_p^{l},v_p^{l-1}} = {d^{-\frac{1}{2}}_{v_p^{l}}}{d^{-\frac{1}{2}}_{v_p^{l-1}}}$ is the normalized weight of the edge $(v_p^{l},v_p^{l-1})$ in the path $p$. 
$\Theta = \{ \mathbf{W}^{(l)}\}$ is the $K$-layer GCN model parameters. 
\vspace{-2mm}
\end{restatable}

Then, according to 
Equation (\ref{derivate_h_v_k}) in Lemma~\ref{lemma:gcn_influence}, 
the target node $v$'s final node representation 
$\tilde{\mathbf{h}}_v^{(K)}$ learnt on the perturbed graph $\tilde{G}$
can be expressed as 
$\tilde{\mathbf{h}}_v^{(K)} = \sum_{u\in \tilde{\mathbf{\Lambda}}_v^{(K)}}\frac{\partial \tilde{\mathbf{h}}_v^{(K)}}{\partial \mathbf{h}_u^{(0)}} \cdot \mathbf{h}_u^{(0)}$, 
where $\tilde{\mathbf{\Lambda}}_v^{(K)}$ is the node set containing 
$v$'s neighbors within $K$-hop on the perturbed graph $\tilde{G}$, i.e., after modifying the edge status between the target node $v$ and  other nodes in the clean graph $G$.

Then, 
the 
attack's objective function in Equation (\ref{eqn:attack}) is equivalent to the following objective function:
{
\small
\begin{align}
\label{obj_h_der}
    & \max_{\tilde{\mathbf{A}}_v} \big([\sum_{u\in \tilde{\mathbf{\Lambda}}_v^{(K)}}\frac{\partial \tilde{\mathbf{h}}_v^{(K)}}{\partial \mathbf{h}_u^{(0)}} \cdot \mathbf{h}_u^{(0)}]_c - [\sum_{u\in \tilde{\mathbf{\Lambda}}_v^{(K)}}\frac{\partial \tilde{\mathbf{h}}_v^{(K)}}{\partial \mathbf{h}_u^{(0)}} \cdot \mathbf{h}_u^{(0)}]_{y_v}\big) \notag \\
    & \, s.t., \quad \sum\nolimits_{s} |\tilde{A}_{v,s} - A_{v,s}| \leq \Delta,
\end{align}
%\end{equation}
}%

Finally, based on the following Assumption~\ref{assump:fea_indep} and Theorem~\ref{thm:relate}, we reach Theorem~\ref{thm:attack_label_inf}
that reformulates the  evasion attack's objective function via label influence. 
We also conduct experiments (See Section~\ref{eval_results}) to verify that Assumption~\ref{assump:fea_indep} holds in practice. 

\begin{restatable}[]{assump}{assumption} 
\label{assump:fea_indep} 
Given a target node $v$ and a target label $c$. 
We assume that any node $u$, within the $K$-hop neighbor of $v$, has a negligible feature-label influence on $v$ if $u$ is not a label-$c$ node. 
Formally, 
{
\begin{equation}
\label{feature_independence}
    [\frac{\partial \tilde{\mathbf{h}}_v^{(K)}}{\partial \mathbf{h}_u^{(0)}} \cdot \mathbf{h}_u^{(0)} ]_c \approx 0, \quad 
    \forall \, u \in \tilde{\mathbf{\Lambda}}_v^{(K)},\ y_u \neq c.
\end{equation}
}%
\end{restatable}

\begin{theorem}
\label{thm:attack_label_inf}
Let $\tilde{I}_{l}(v,u;K)$ be the label influence of node $u$ on the target node $v$ with $K$ iterations of LP after the attack. Then, the attack's objective function in Equation (\ref{eqn:attack}) equals to the following objective function 
on label influence:
{
\small
\begin{align}
    & \max_{\tilde{\mathbf{A}}_v} \big (\sum_{\substack{u\in \tilde{\mathbf{\Lambda}}_v^{(K)}, y_u=c}}\tilde{I}_{l}(v,u;K) 
    -\sum_{\substack{z\in \tilde{\mathbf{\Lambda}}_v^{(K)} , y_z=y_v}}\tilde{I}_{l}(v,z;K) \big), \notag \\ 
    & \, s.t., \quad \sum\nolimits_{s} |\tilde{A}_{v,s} - A_{v,s}| \leq \Delta, \label{obj:attack_label_inf}
\end{align}
}%
where $\tilde{I}_{l}(v,u;K)$ 
is defined as:
{
\small
%\footnotesize
\begin{align}
    \label{eqn:norm_weight}
    \tilde{I}_{l}(v,u;K) = \sum_{p=1}^{\tilde{\varPsi}_{v \rightarrow u}}
    \prod_{l=K}^{1} \tilde{d}^{-\frac{1}{2}}_{v_p^{l}} \tilde{d}^{-\frac{1}{2}}_{v_p^{l-1}},
\end{align}
}%
where $\tilde{\varPsi}_{v \rightarrow u}$ is the total number of paths $[v_p^K,v_p^{K-1},$ $\cdots,v_p^{1},v_p^{0}]$ of length $K+1$ from $v$ to $u$ on the perturbed graph $\tilde{G}$, where $v_p^K=v$ and $v_p^0=u$. $\tilde{d}_u$ is $u$'s degree 
on 
the perturbed graph 
$\tilde{G}$ 
and 
${\tilde{d}^{-\frac{1}{2}}_{v_p^{l}}}{\tilde{d}^{-\frac{1}{2}}_{v_p^{l-1}}}$ is the normalized weight of the edge $(v_p^{l},v_p^{l-1})$ in path $p$ in 
$\tilde{G}$. 
\vspace{-2mm}
\end{theorem}

We have the following observations from Theorem~\ref{thm:attack_label_inf}.
\begin{itemize}
\vspace{-1mm}
\item Our attack does not need to operate on model parameters $\Theta^*$, different from existing attacks that involve dense multiplication on $\Theta^*$. Thus, our attack is more efficient. 

\item Our attack can be applied to any layer 
GNN, as the label influence is defined for general $K$-iteration LP. 
However, most of the existing attacks can only directly attack two-layer GNNs.
Thus, our attack is more practical. 

\item 
The only information our attack needs 
to know 
is
 the target node $v$'s within $K$-hop neighbors, whose labels are $y_v$ or $c$.  
In practice, if the labels of these node are unknown, we can estimate 
them via querying the GNN model, and treat the estimated labels as the true labels. Thus, our attack can be seen a restricted black-box attack. 

\end{itemize}

Next, we show how to fast calculate the label influence and design our influence-based targeted evasion attack.

\subsection{Efficient Calculation of Label Influence} 

According to Theorem~\ref{thm:attack_label_inf}, 
the attack's goal is to select the minimum set of nodes such that when changing the edge status between the target node $v$ and these selected nodes,
the difference between the two label influence terms will be maximized.  
Observing Equation (\ref{obj:attack_label_inf}), we note that the two label influence terms are defined on two sets of nodes:  a set of nodes 
having the same label  as the target label $c$, and a set of nodes 
having the same label as the target node's label $y_v$. 
Intuitively, if we add an edge between $v$ and a label-$c$ node, we can make $v$ be close to label $c$; and 
if we remove an edge between $v$ and a label-$y_v$ node, we can make $v$ away from label $y_v$. 
Thus, our 
idea to solve 
Equation (\ref{obj:attack_label_inf}) is as follows:  
\begin{itemize}[leftmargin=*]
\vspace{-1mm}
\item First, we define a candidate set $\mathcal{N}_A \subset \{ y_u=c, u \in{\mathbf{\Lambda}}_v^{(K)}\}$ 
which contains label-$c$ nodes 
that are 
\emph{not} connected with $v$ in the clean graph, as well as 
a candidate set $\mathcal{N}_B \subset \{ y_z= y_v, z \in {\mathbf{\Lambda}}_v^{(K)}\}$ 
which contains label-$y_v$ nodes that are connected with $v$ in the clean graph. 
We denote $\mathcal{S}$ as the final selected nodes from $\mathcal{N}_A$ and $\mathcal{N}_B$, and initialize $\mathcal{S} = \{ \}$. 
For each node $u \in \mathcal{N}_A \cup \mathcal{N}_B \setminus \mathcal{S}$, we change the edge status between $v$ and $u$ and compute the gap between two label influence terms.  
\item Next, we record the node $u^*$ that obtains the largest positive gap. 
Then, we modify the edge status between $v$ and $u^*$, calculate the value of the attack's objective function, and update $\mathcal{S} = \mathcal{S} \cup \{u^*\} $. 
\item We repeat above steps at most $\Delta$ times and break if the value of attack's objective function is bigger than 0. 
Finally, we have the attack edges $\{(v, u^*), u^* \in \mathcal{S}\}$. 
\end{itemize}

However, note that when modifying the edge status 
between $v$ and $u^*$, 
the normalized weight % $\tilde{a}_{v_p^{l},v_p^{l-1}}$ 
%${\tilde{d}^{-\frac{1}{2}}_{v_p^{l}}}{\tilde{d}^{-\frac{1}{2}}_{v_p^{l-1}}}$
for all edges containing $u^*$
% $(v_p^{l},v_p^{l-1})$ 
in all paths $\tilde{\varPsi}_{v \rightarrow u}$ in Equation (\ref{eqn:norm_weight}) should be recalculated. 
When the candidate set has a large size or/and the  number of recalculated edge weights is large, calculating the exact label influence will have a large computational complexity.
To solve the problem, we propose an approximate algorithm to efficiently compute the label influences.  
More details are in 
Supplementary Material.

Algorithm~\ref{alg:cal_LI} in the full report illustrates how we efficiently calculate the label influences via depth first search (DFS), and Algorithm~\ref{alg:our_attack} in the report shows the details of implementing our attack. %, respectively. 

\begin{figure*}[!t]
\vspace{-2mm}
\center
\subfigure[Cora]{\includegraphics[width=0.28\textwidth]{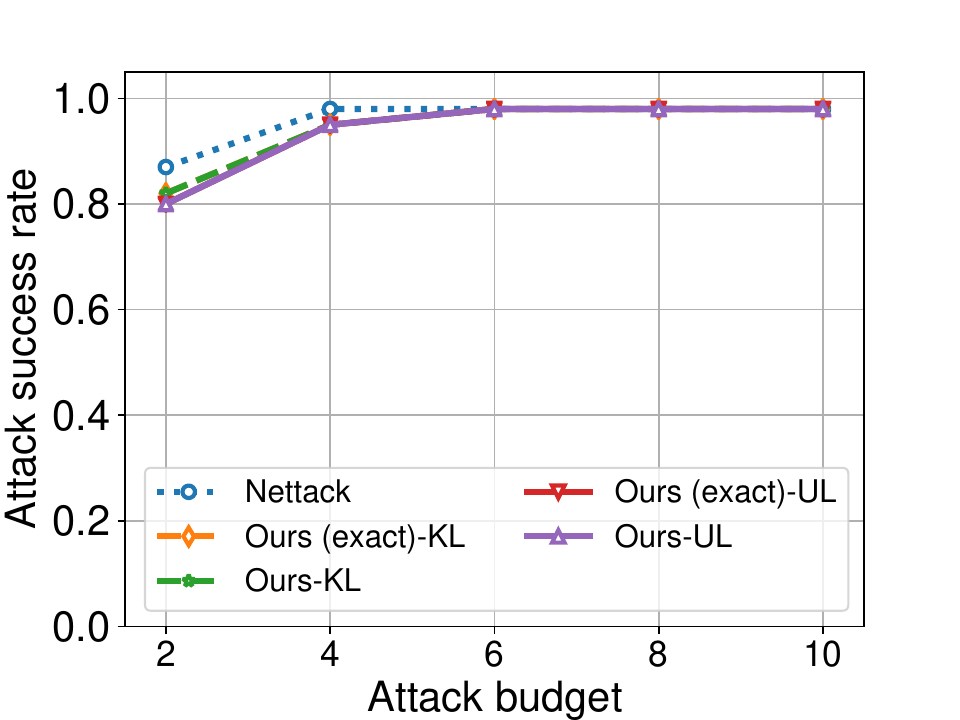}}
\subfigure[Citeseer]{\includegraphics[width=0.28\textwidth]{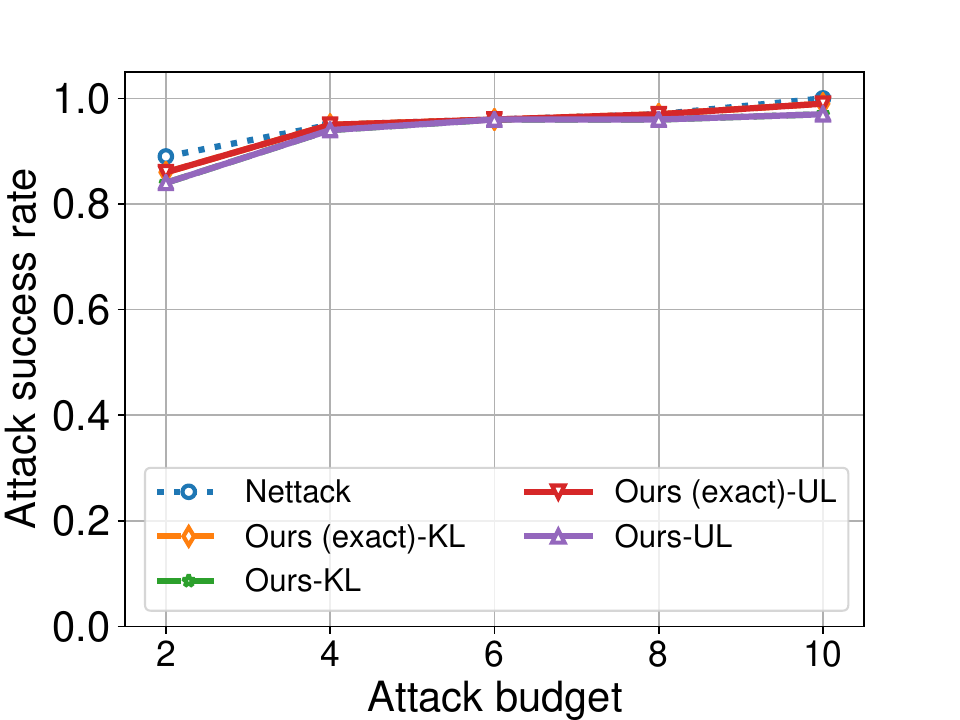}}
\subfigure[Pubmed]{\includegraphics[width=0.28\textwidth]{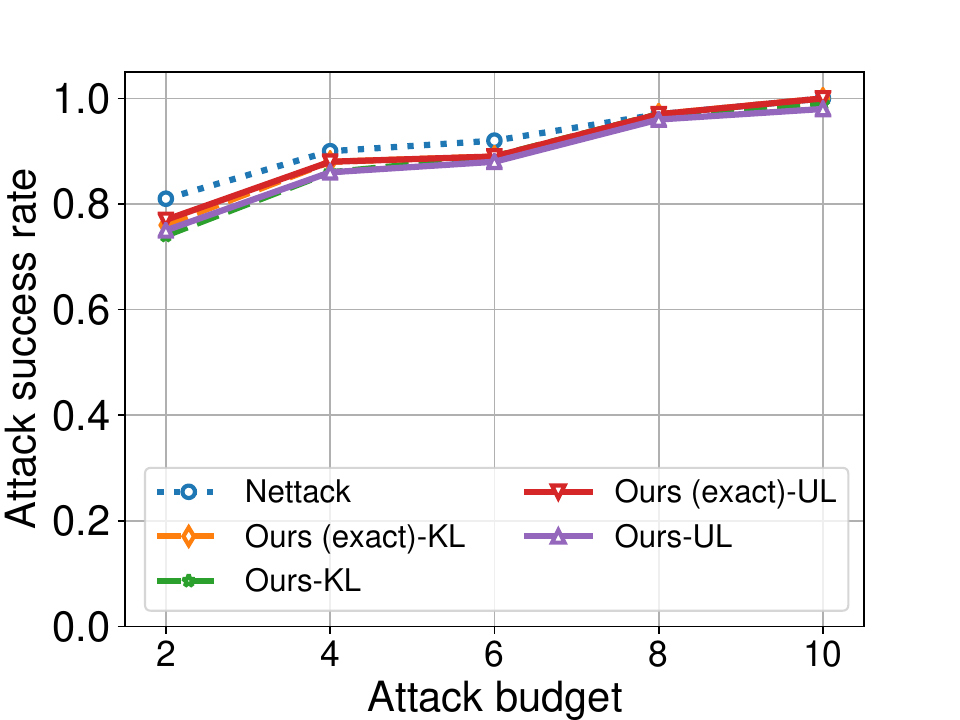}}
\vspace{-4mm}
\caption{Attack success rate vs. attack budget per target node on a two-layer GCN of all compared attacks on the three graphs. }
\label{fig:attack_2layer_GCN}
\vspace{-4mm}
\end{figure*}

\section{Evaluation}
\label{sec:eval}

\subsection{Experimental Setup}

\noindent  {\bf Datasets.} Following 
\cite{zugner2018adversarial,xu2019topology,geisler2021robustness}, we use three benchmark graphs (i.e., Cora, Citeseer, and Pubmed)~\cite{sen2008collective} to evaluate our attack. 
In these graphs, each node represents a documents and each edge indicates a citation between two documents.  
Each document treats the bag-of-words feature as the node feature vector, and has a label. 
Table~\ref{dataset_stat} %in Appendix 
in the full version 
shows basic statistics of these 
graphs. 

\noindent  {\bf Training nodes and target nodes.}
We use the public training nodes to train GNN models, and 
target nodes to evaluate attacks against the trained GNN models. 
For the target nodes, we employ a random sampling technique to select 100 nodes that are correctly classified by each GNN model as the target nodes.
Similar to Nettack~\cite{zugner2018adversarial}, for each target node, we choose the predicted label by the GNN model with a second largest probability as the target label.

\vspace{+0.5mm}  \noindent  {\bf Compared attacks.} 
We compare our influence-based attack with the state-of-the-art Nettack~\cite{zugner2018adversarial} for attacking two particular GNNs: GCN and SGC. 
Note that Nettack is mathematically designed to only attack two-layer GNNs and cannot 
directly attack multi-layer GNNs. To attack multi-layer GNNs, Nettack needs to be performed via an indirect way: It first attacks a surrogate two-layer GNN to generate the attack edges, and then 
transfers these attack edges to attack the target  multi-layer  GNNs. 
When computing the label influence, our attack needs to know the labels of unlabeled nodes in the graph. 
When our attack knows the the true labels, we denote it as \textbf{Ours-KL}. 
When the true labels are unknown, our attack first queries the learnt 
GNN model to estimate labels for unlabeled nodes and then uses the estimated labels as the true labels. We denote this variant as \textbf{Ours-UL}. 
As a comparison, we also test our attack that is implemented based on exact label influence calculation, and denote the corresponding two methods with known and unknowns labels as \textbf{Ours (exact)-KL} and \textbf{Ours (exact)-UL}, respectively. 

\vspace{+0.5mm}  \noindent  {\bf Evaluation metric.} For graph perturbation attacks, we adopt attack success rate and running time as the metrics. 
 Given a target GNN model, a set of target nodes, target label, and an attack budget $\Delta$, attack success rate is the fraction of target nodes that are misclassified by the target GNN to be the target label when the number of attack edges per target node is at most $\Delta$. Running time is reported on average across all the target nodes.

\vspace{+0.5mm}  \noindent  {\bf Implementation.} 
We train all GNNs using the public
source code. 
We test Nettack using the source code (\url{https://github.com/danielzuegner/nettack}). 
We implement our attack in PyTorch. 
All experiments are conducted on an A6000 GPU with 48G memory.
\emph{Due to space limitation, we only show comparison results with GCN, and all  results are in the full version: \url{https://github.com/ventr1c/InfAttack}.}

\begin{figure*}[!t]
\vspace{-2mm}
\center
\subfigure[Cora]{\includegraphics[width=0.28\textwidth]{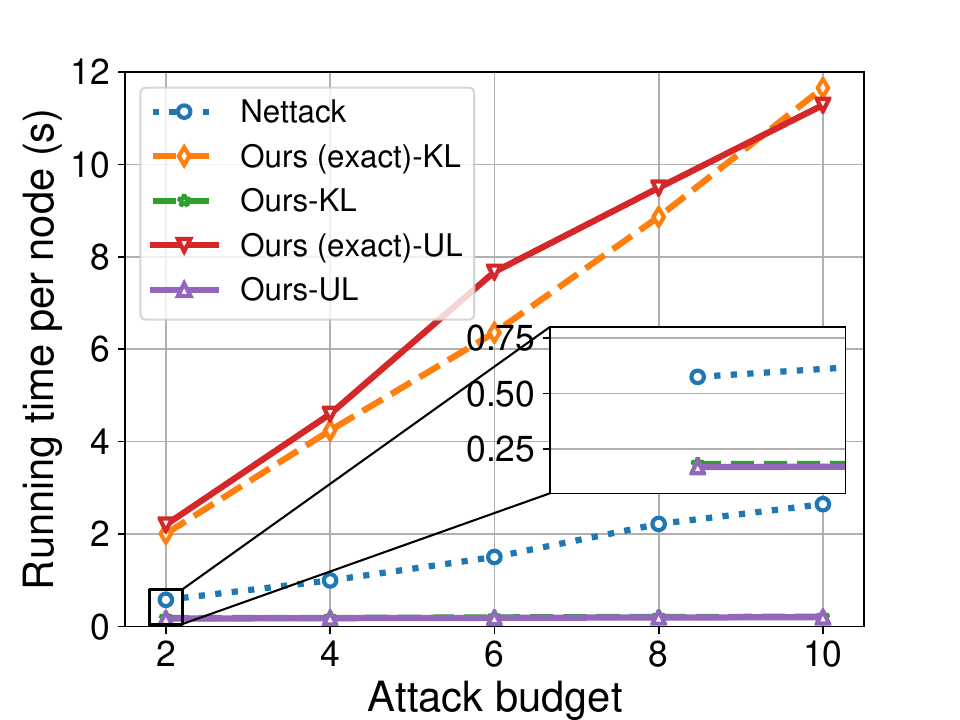}}
\subfigure[Citeseer]{\includegraphics[width=0.28\textwidth]{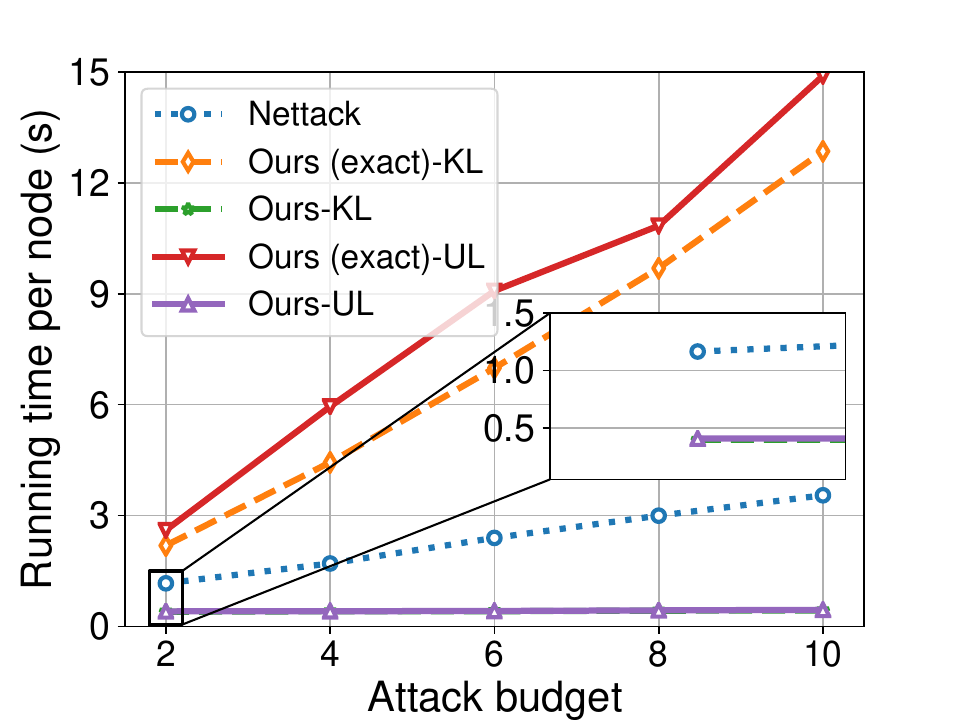}}
\subfigure[Pubmed]{\includegraphics[width=0.28\textwidth]{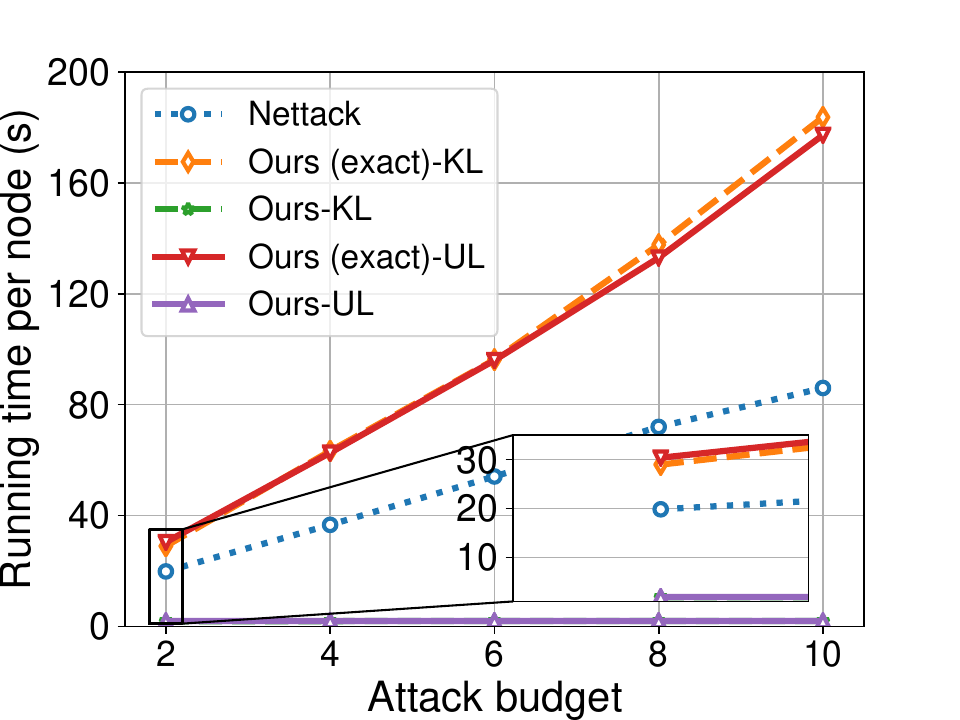}}
\vspace{-4mm}
\caption{Running time vs. attack budget per target node on two-layer GCN of all compared attacks on the three graphs.}
\label{fig:time_2layer_GCN}
\vspace{-2mm}
\end{figure*}

\subsection{Results on Attacking Two-layer GNNs}
\label{eval_results}

\vspace{+0.5mm}  \noindent  {\bf Results on attacking two-layer GCN/SGC.}
We compare all attacks 
in terms of effectiveness (i.e., attack success rate) and efficiency (i.e., running time) against two-layer GCN/SGC. 
Figure~\ref{fig:attack_2layer_GCN} and Figure~\ref{fig:attack_2layer_SGC} in the full version 
show the attack success rate  
against GCN and SGC on the three graphs, respectively.
Moreover, Figure~\ref{fig:time_2layer_GCN} and Figure~
\ref{fig:time_2layer_SGC} 
in the full version 
show the running time of all attacks 
against GCN and SGC on the three graphs, respectively. 
We have the following key observations. 
\begin{itemize}[leftmargin=*]
\item \emph{Our attacks based on approximate label influence  have similar performance with those based on exact label influence, but is much more efficient.} Specifically, the difference of the attack success rate between the two is less than $2\%$ in all cases.
This shows that our proposed efficient algorithm for label influence  calculation is effective enough. 
Moreover, our attacks based on approximate label influence 
are 1-2 orders of magnitude more efficient than those based on exact label influence. 
\item \emph{Our attacks with true labels and with estimated labels have similar performance.} Specifically, the difference of the attack success rate between Ours-KL and Our-UL is negligible, i.e., less than $2\%$ in all cases, and the running time of both Ours-KL and Our-UL are almost the same. 
One reason is that the trained GNN model has accurate predictions on the unlabeled nodes, and thus most of the estimated labels match the true labels. 
One should note that Ours-UL knows very limited knowledge about the GNN model and thus it is a very practical attack. 

\item \emph{Our attacks achieve comparable performance with Nettack}. 
Nettack achieves state-of-the-art attack performance against two-layer GCN. 
Our attacks have a slightly lower attack success rate than Nettack when the attack budget is small, e.g., less than 4. 
This is possibly because our attack use some approximations on Assumption~\ref{assump:fea_indep}, and when the attack space is small, Assumption~\ref{assump:fea_indep}   negatively affects the attack effectiveness to some extent.  
However, when the attack budget is larger than 4, our attacks obtain almost the same performance with Nettack. 
 
\item \emph{Our attacks are much more efficient than Nettack}. Specifically, our attacks have a 5-50x speedup over Nettack across the three graphs. As the attack budget increases (from 2 to 10) or the graph size increases (from Cora to Pubmed), our attacks achieve better efficiencies.
The reasons are two-fold. 
First, Nettack needs to multiply GNN model parameters in different layers, while our attacks do not.   
Second, Nettack involves multiplying the node hidden features, while ours is performed by calculating the label influence. Node hidden features are often high-dimensional, while label influence only needs scalar edge weights products.
\end{itemize}

\begin{figure}[!t]
  \centering
  \includegraphics[width=0.3\textwidth]{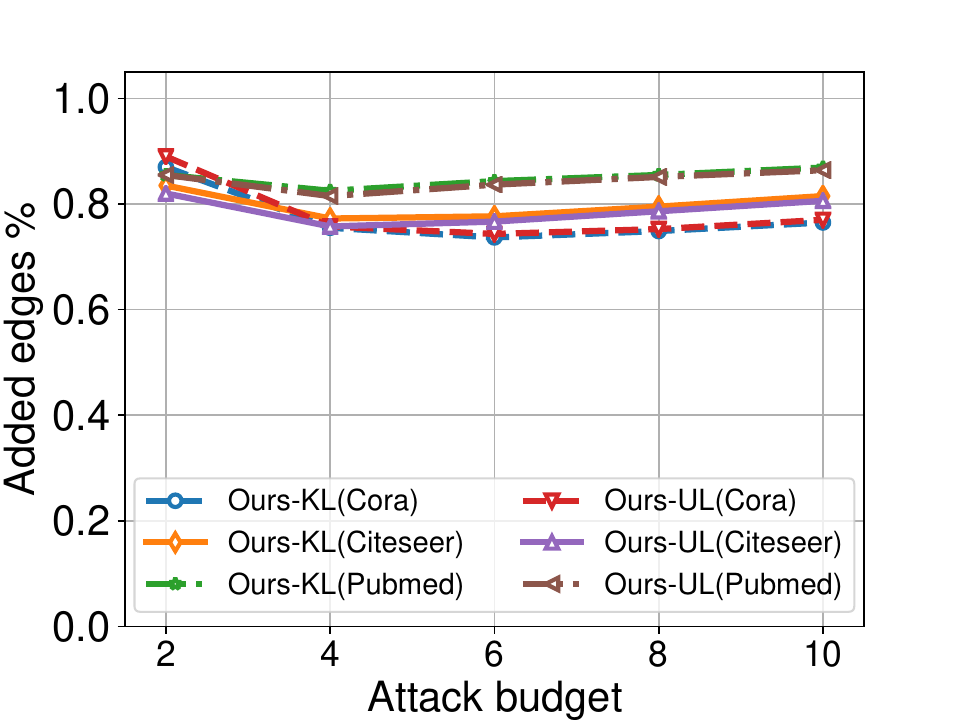} 
    \vspace{-2mm}
    \caption{Fraction of the added edges among all attack edges generated by our attacks against a two-layer GCN vs. attack budget 
    per target node.} 
    \label{fig:Added_edges_GCN}
   \vspace{-2mm}
\end{figure}

\vspace{+0.5mm}  \noindent  {\bf Analysis of the attack edges.}
We further analyze the properties of the attack edges.
Figure~\ref{fig:Added_edges_GCN} and Figure~\ref{fig:Added_edges_SGC} in the full version 
show the fraction of the added edges generated by our attacks against two-layer GCN and two-layer SGC, respectively. 
First, Ours-KL and Ours-UL generate almost the same fraction of added edges in all attack budgets and all graphs. This again verifies the similar characteristics between Ours-KL and Ours-UL. 
Second, the fraction of added edges is larger than 0.5 in all cases. This indicates
that when performing the targeted attack,  
adding new edges between the target node and the nodes with the target label could be more effective than removing existing edges between the target node and the nodes having the same label as the target node. 

\vspace{+0.5mm}  \noindent  {\bf Analysis of the factors that affect the attack performance.} 
We consider the following three factors, i.e., \emph{node degree}, \emph{node centrality}, and \emph{graph size}, that could affect the target node's attack performance. Here, we  adopt the normalized closeness centrality (NCC) as the metric to measure node centrality.  Specifically, the NCC of a node is the average length of the shortest path between the node and all other nodes in the graph. 
We have the following conclusions: 
\textbf{1)} \emph{Nodes with smaller degrees are easier to attack}. Given a fixed attack budget (e.g., 4 in our experiment), we observe that in all the three datasets, 
100\% of the target nodes with degree $<=4$ attack successfully, while at most 83\% and 85\% of the target nodes with degree $>4$ attack successfully against 2-layer GCN and 2-layer SGC, respectively. 
\textbf{2)} \emph{Nodes with larger centrality are easier to attack.} 
We assume the attack budget is 4 per node. 
Specifically, 100\% and 98\% of the 50 target nodes with the largest NCC successfully attack 2-layer GCN and 2-layer SGC in the three datasets, while $<90\%$ and $<80\%$ of the 50 target nodes with the smallest NCC can successfully perform the attack. 
\textbf{3)} \emph{No obvious relationship between graph size and attack success rate}. Specifically, 
graph size: 
Pubmed $>$ Citeseer $>$ Cora. When attacking 2-layer GCN and the attack budget is 6, we have the attack success rate: Cora (0.98) $>$ Citeseer (0.96) $>$ Pubmed (0.87). 
However, when the attack budget is 10, we have the attack success rate: Pubmed (1.00) $>$ Citeseer (0.99) $>$ Cora (0.98).

\begin{figure}[t]
\center
\subfigure[Attack success rate]{\includegraphics[width=0.225\textwidth]{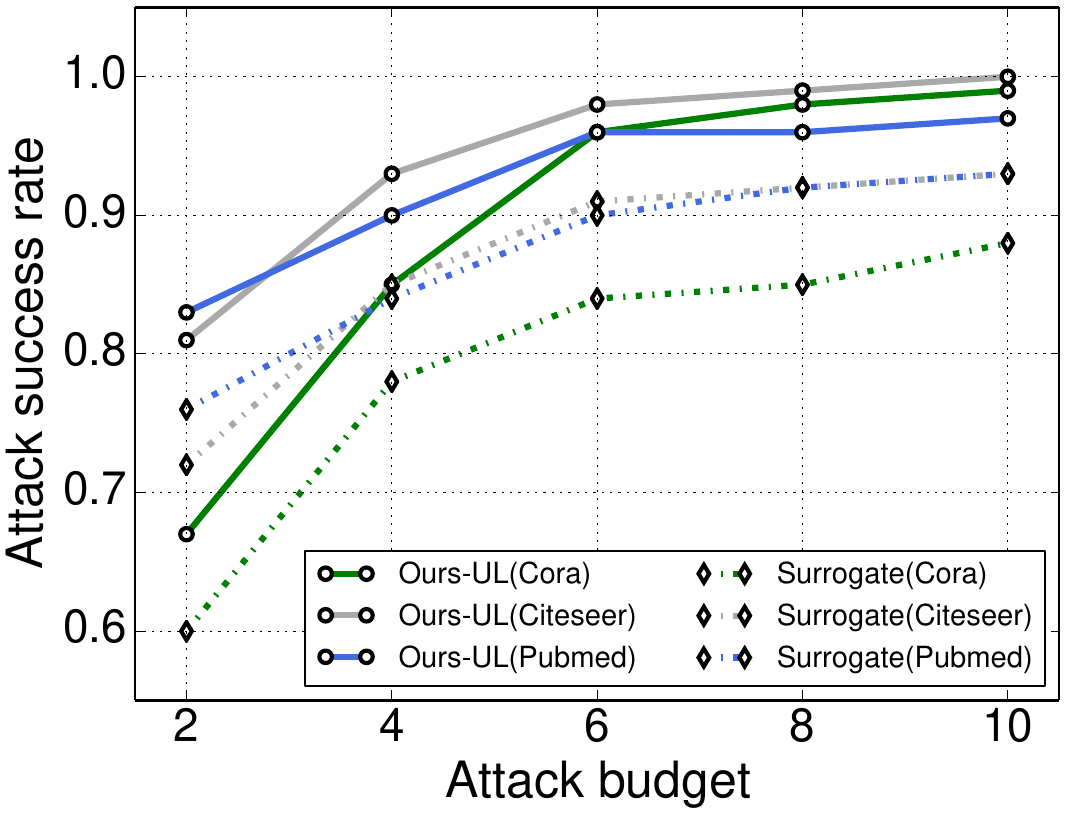} \label{fig:GCN_4layer_attack}}
\subfigure[Running time]{\includegraphics[width=0.225\textwidth]{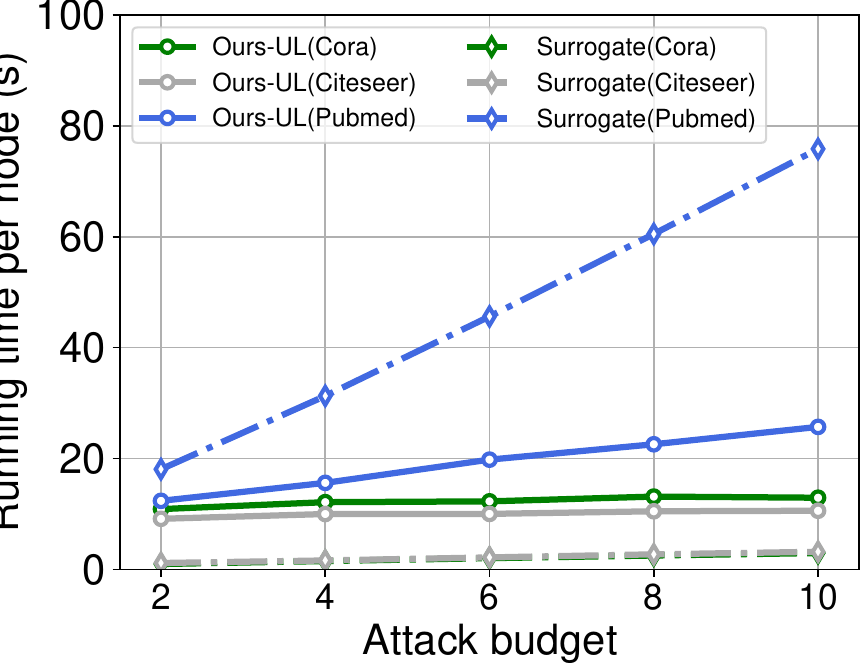}\label{fig:GCN_4layer_time}}
\vspace{-4mm}
\caption{(a) Attack success rate and (b) Running time of our attacks against four-layer GCN on the three graphs. }
\vspace{-4mm}
\end{figure}

\subsection{Results on Attacking Multi-layer GNNs}

\vspace{+0.5mm}  \noindent  {\bf Attack performance on four-layer GCN/SGC.}
In this experiment, we evaluate our attacks against multi-layer GCN/SGC. 
We denote Nettack that attacks a surrogate two-layer GNN model first and then transfers to attacking the target model as {\bf Surrogate}. 
Figure~\ref{fig:GCN_4layer_attack} 
and Figure~\ref{fig:SGC_4layer_attack} 
in the full version 
show the attack success rate of our attack vs. attack budget against four-layer GCN and four-layer SGC on the three graphs, respectively. 
First, similarly, 
our attacks with both known label and unknown label are effective and achieve close attack performance, and we thus show results with unknown label for simplicity. 
When the attack budget 
is 6, our attacks achieve an attack success rate of 
$\geq 90\%$ in all cases. 
Second, our attacks are more effective than the \emph{indirect} surrogate model based attacks. Specifically, our attacks have more than $10\%$ higher attack success rate than the surrogate model based attacks in almost all cases.  

Moreover, Figure~\ref{fig:GCN_4layer_time} 
and Figure~\ref{fig:SGC_4layer_time} in the full version  
shows the running time of our attacks vs. attack budget against four-layer GCN and four-layer SGC on the three graphs, respectively. 
Our attack is efficient. 
For instance, it takes our attacks less than
$25s$ on average to attack a target node on the largest Pubmed in all cases. However,  
the surrogate model costs about $85s$, validating that our attack is much more efficient. 

\vspace{+0.5mm}  \noindent  {\bf Transferring our attack to other GNNs.}
In this experiment, we study the transferability of 
our attacks, i.e., whether the attack edges generated by our attacks against GCN/SGC can be also effective for other GNNs. 
Specifically, we use our attacks to generate the attack edges for each target node by attacking the source GNN (GCN or SGC), change the graph structure based on the attack edges, and adopt a target GNN to classify each target node on the perturbed graph. 
We select two additional representative GNNs, i.e., GAT~\cite{velivckovic2018graph} and JK-Net~\cite{xu2018representation}, as the target GNN. If a target node is also misclassified by the target GNN to be the target label, we say the attack edges generated by the source GNN are transferable.

Table~\ref{tbl:transfer_2layer_GCN} and Table~\ref{tbl:transfer_2layer_SGC} in the full version 
show the attack success rate of transferring of our attacks against two-layer GCN and two-layer SGC to attack other GNNs on the three graphs, where the attack budget per target node is 6. 
Note that we also show the attack performance for target GNNs without attack, i.e., the prediction error of target GNNs on the target nodes in the clean graph. 
We have the observations:  
First, our attacks against GCN (or SGC) have the best transferability to SGC (or GCN).
This is because SGC is a special case of GCN and they share similar model architectures. 
Second, our attacks are also effective against GAT and JK-Net. 
Specifically, on all the three graphs, our attacks can increase the classification errors by at least $60\%$ when attacking GAT and JK-Net. 
This indicates that all the attack edges generated by our attack on the source GNN can be transferred to attack the target GNNs. Such good transferability further demonstrates the advantages of using (label) influence to perform the target evasion attacks.

\begin{table}[!t]\renewcommand{\arraystretch}{1.1}
\centering
\caption{Transferability of our attacks against two-layer GCN to other GNNs. Attack budget per target node is 6.}
\small
\begin{tabular}{|c|c|c|c|c|c|}
\hline
{\bf Dataset}   & {\bf Source}   & \multicolumn{4}{c|}{\bf Target} \\ \hline

\multirow{4}{*}{\bf Cora} & {\bf GCN} &  {\bf GCN}   &  {\bf SGC}   &    {\bf GAT} & {\bf JK-Net}    \\ \cline{2-6} 
&  {\bf No attack}   &  0   &  0.01   &  0.03   & 0.02     \\ \cline{2-6} 
&  {\bf Ours-KL}   &  0.98   &  0.82   &  0.66   & 0.67     \\ \cline{2-6} 
& {\bf Ours-UL}    &  0.98 & 0.84     & 0.65    & 0.70 \\ \hline \hline

\multirow{4}{*}{\bf Citeseer} & {\bf GCN} &  {\bf GCN}   &  {\bf SGC}   &    {\bf GAT} & {\bf JK-Net}    \\ \cline{2-6} 
&  {\bf No attack}   &  0   &  0.01   &  0.01   & 0.03     \\ \cline{2-6}
&  {\bf Ours-KL}  & 0.96    &  0.78   &    0.70 &  0.63   \\ \cline{2-6} 
&  {\bf Ours-UL} & 0.96    &  0.78     & 0.72    & 0.63  \\ \hline \hline

\multirow{4}{*}{\bf Pubmed} & {\bf GCN} &  {\bf GCN}   &  {\bf SGC}   &    {\bf GAT} & {\bf JK-Net}    \\ \cline{2-6} 
&  {\bf No attack}   &  0   &  0.03   &  0.04   & 0.05     \\ \cline{2-6}
& {\bf Ours-KL}    &  0.89   & 0.80    & 0.80 & 0.79    \\ \cline{2-6} 
& {\bf Ours-UL}  & 0.88    & 0.80      & 0.80   & 0.77 \\ \hline
\end{tabular}
\label{tbl:transfer_2layer_GCN}
\end{table}

\begin{table}[!t]\renewcommand{\arraystretch}{1.0}
\centering
\small
%\vspace{+1mm}
\caption{Attack results of Ours-KL on OGB-arxiv.}
% \addtolength{\tabcolsep}{-2.5pt}
% \vspace{-2mm}
\begin{tabular}{|c|c|c|c|c|c|c|}
\hline
 {\bf Attack budget}                 &  \bf 2     &   \bf  4   &  \bf 6     & \bf 8      &   \bf 10    &  \bf  Time    \\ \hline
     {\bf Nettack} & OOM & OOM & OOM & OOM & OOM & -  \\ \hline
     {\bf Ours-GCN}             &   0.65    &   0.72    &  0.73     &  0.73    &  0.82     &   40.1s   \\ \hline
     {\bf Ours-SGC}             &   0.90    &   0.93    &  0.95     &  0.95    &  0.96     &   40.7s   \\ \hline
\end{tabular}
\label{tbl:largescale_atk_results}
\vspace{-4mm}
%\vspace{-1.5em}
\end{table}

% \vspace{-2mm}
\section{Discussions}
%\subsection{Comparing with More Attack Baselines}

\vspace{+0.5mm} \noindent {\bf Attack performance on a large-scale dataset.} 
We conduct experiments on the large-scale dataset OGB-arxiv~\cite{hu2020open} to demonstrate the superior efficiency of our proposed attack method compared to baselines. We use a $2$-layer GCN/SGC as the target GNN, achieving a clean accuracy of approximately 60\% on test nodes. 
The attack method is set as {Ours-KL}. 
Note that time is denoted as the average running time of compared attack methods across the five attack budgets. 
We assess the attack performance on 100 target nodes that are accurately classified by GCN/SGC. The comparison results between Nettack and our attack are presented in Table~\ref{tbl:largescale_atk_results}. We observe that \textbf{1)} Nettack encounters an out-of-memory (OOM) error on our platform due to the need for storing dense model weights and involving intensive matrix-matrix multiplication. \textbf{2)} Our attacks achieve highly promising attack success rates while maintaining efficiency. Specifically, with an attack budget of $6$, the attack success rates of {Ours-KL} against GCN and SGC are $73$\% and $95$\%, respectively. These results demonstrate the significant advantages of our attack method over baselines on large graphs.

\noindent {\bf Comparing with more attack baselines.}
To further demonstrate the effectiveness of our attack, 
we compare Ours-KL with a more recent attack IG-FGSM~\cite{wu2019adversarial}, where we use the same setting as that in Section~\ref{eval_results} (i.e., 2-layer GCN/SGC as the target GNN, 100 target nodes). The comparison results on Cora are reported in Table~\ref{tbl:compared_with_igfgsm_gcn} and Table~\ref{tbl:compared_with_igfgsm_sgc} in the full report. 
Note that time is denoted as the average running time of compared attack methods across the five attack budgets. 
We observe that: \textbf{1)} Nettack not only outperforms IG-FGSM when the attack budget is small, but also is far more efficient than IG-FGSM. Specifically, when the attack budget is $2$, the attack success rates of IG-FGSM agasint GCN and SGC are only $0.29$ and $0.32$, respectively, which are significantly lower than that of Nettack, $0.85$ and $0.65$. Moreover, the running times of IG-FGSM against GCN and SGC are $62$s and $55$s, respectively, which are much lower than that of Nettack, $1.5$s and $2.5$s, respectively. \textbf{2)} Our method is even more efficient than Nettack. Specifically, when attacking GCN and SGC, the running time of our method is only $0.1$s, which is lower than other two methods, further validating the efficiency of our method.

\begin{table}[!t]\renewcommand{\arraystretch}{1.0}
\centering
\small
\caption{Comparing our attack vs. IG-FGSM~\cite{wu2019adversarial} on Cora.}
\begin{tabular}{|c|c|c|c|c|c|c|c|}
\hline
{\bf Model} & {\bf budget}                 &  \bf 2     &   \bf  4   &  \bf 6     & \bf 8      &   \bf 10    &  \bf  Time    \\ \hline
\multirow{3}{*}{\bf GCN} & {\bf IG-FGSM} & 0.29 & 0.75 & 0.89 & 0.92 & 0.94 & 62s  \\ \cline{2-8}
   &  {\bf Nettack}             &   0.85    &   0.96    &  0.97     &  0.97    &  0.97     &   1.5s   \\ \cline{2-8}
   &  {\bf Ours}             &   0.90    &   0.93    &  0.95     &  0.95    &  0.96     &   0.1s   \\ \hline 
\end{tabular}
\label{tbl:compared_with_igfgsm_gcn}
\vspace{-2mm}
\end{table}

\begin{table}[t]\renewcommand{\arraystretch}{1.0}
\centering
\caption{
Comparing our attack vs. black-box attack \cite{wang2022bandits}. 
}
\small
\centering
\addtolength{\tabcolsep}{-2pt}
\begin{tabular}{|c|c|c|c|c||c|c|c|c|c|} \hline 

{\bf GCN} & {\bf budget}  & {\bf 2} & {\bf 4} & {\bf 6}  & {\bf GCN}
 & 
{\bf budget}  & {\bf 2} & {\bf 4} & {\bf 6} \\ \hline
\multirow{2}{*}{\bf Cora}  & {\bf Ours}  &  {0.81} & {0.94} & {0.98} & \multirow{2}{*}{\bf Citeseer} & {\bf Ours}  &  {0.86} & {0.94} & {0.95}  \\ \cline{2-5} \cline{7-10}
& {\bf \cite{wang2022bandits}}  &  {0.68} & {0.79} & {0.80}  & 
& {\bf \cite{wang2022bandits}}  &  {0.84} & {0.91} & {0.92} \\ 
\hline 
\end{tabular}
\label{tbl:res_BBattack_gcn}
\vspace{-4mm}
\end{table}

\noindent {\bf Comparing with black-box attacks.}
In the paper, we mainly compare our attack with the white-box attack. 
Here we also compare 
with the stringent black-box attacks proposed in~\cite{wang2022bandits}. 
To best explore the attack capability, we do not restrict the number of queries in \cite{wang2022bandits}, and obtain the optimal attack successful rate for a given attack budget.    
Table~\ref{tbl:res_BBattack_gcn} and Table~\ref{tbl:res_BBattack} in the full version show the comparison results on Cora and Citeseer (Note that \cite{wang2022bandits} cannot run on Pubmed due to limited GPU memory) on attacking 2-layer GCN/SGC.  
We can see that our attack is more effective than \cite{wang2022bandits}, especially when the attack budget is small. One key reason is that our attack utilizes the strong connection between GNN and LP, while \cite{wang2022bandits} performs the attack based on the query feedback, i.e., the target node's confidence score after querying the black-box GNN.

\begin{figure}[!t]
  \centering
    \includegraphics[width=0.3\textwidth]{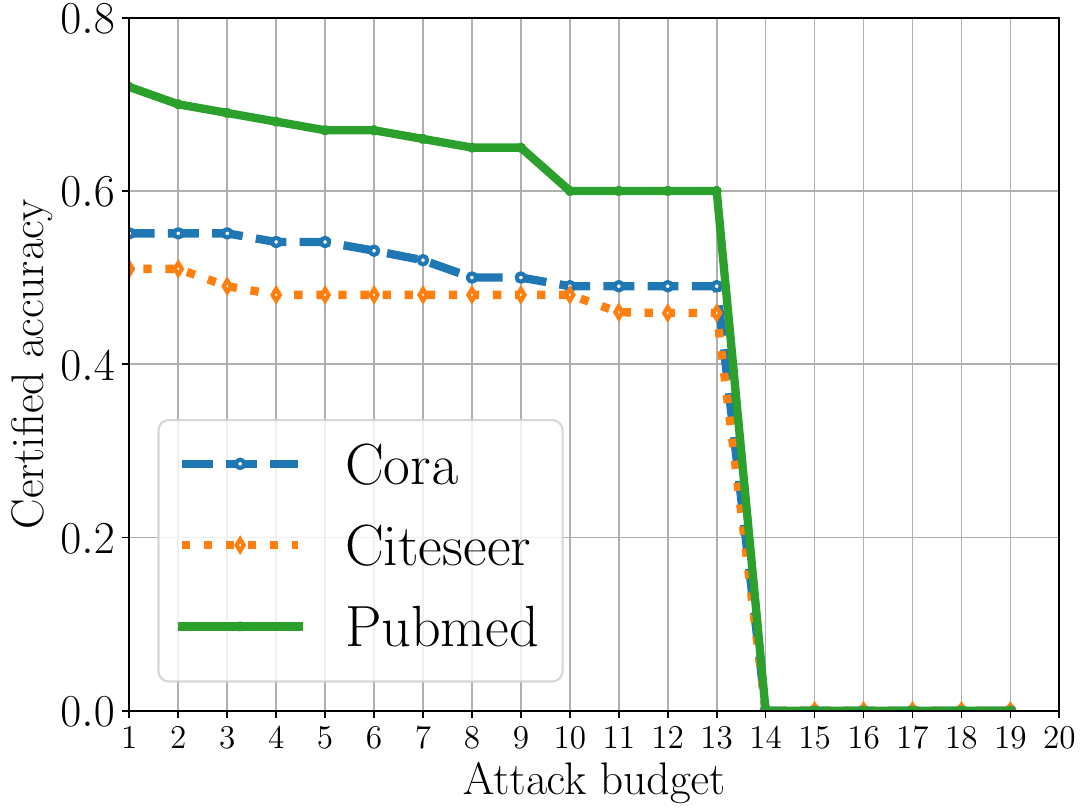}
    \vspace{-2mm}
    \caption{Certified accuracy \cite{wang2021certified} vs. attack budget of GCN.}
    \label{fig:defense}
    \vspace{-6mm}
\end{figure}

\noindent {\bf Defending against our attack.}
As shown in Section~\ref{sec:related}, existing empirical defenses~\cite{zhu2019robust,wu2019adversarial,entezari2020all,zhang2020gnnguard,jin2020graph,geisler2021robustness,zhuang2022defending} are %easy to be 
broken~\cite{mujkanovic2022defenses} when the adversary knows the defense mechanism. Hence, we propose to defend our attack via provable defenses and choose the state-of-the-art randomized smoothing-based provable defense~\cite{wang2021certified}. 
Specifically, given a target node, a model is provably robust for the target node  if the model correctly predicts the same label for the target node when the attacker \emph{arbitrarily} modifies a bounded number of (e.g., at most $R$) edges in the graph, where $R$ is called \emph{certified radius}. 
Hence, provably robust models can defend against the worst-case attack (including our attack). 
Accordingly, certified accuracy under $R$ means the fraction of target nodes that are predicted accurately by modifying any $R$ edges. That is, if a model achieves a larger certified accuracy at a given budget, it shows better provable robustness. We conduct experiments on defending two-layer GCN against the worst-cast attack via \cite{wang2021certified}. Results on the three datasets are shown in Figure~\ref{fig:defense}. We can see that, when any 4 edges are allowed to be modified, the certified accuracy achieved by the method  \cite{wang2021certified} on the three datasets are about 0.50, 0.55, and 0.70, respectively. However, the method \cite{wang2021certified}  cannot provably defend against the worst-case attack when the attack budget is larger than 13, which implies the need to design more powerful provable defenses.      
% \vspace{-2mm}
\section{Conclusion}

We propose an influence-based evasion attack against GNNs. 
Specifically, we first build the connection between GNNs and label propagation (LP) via carefully designed influence functions. 
Next, we reformulate the attack against GNNs to be related to label influence on LP. 
Then, we  design an efficient algorithm to calculate label influences. 
Our attack is applicable to any-layer GNNs and does not need to know the GNN model parameters. 
%We evaluate our attack on multiple benchmark graph datasets. 
Experimental results demonstrate that our attack achieves comparable performance against state-of-the-art white-box attacks, and has a 5-50x speedup when attacking two-layer GCNs. 
Our attack is also effective to attack multi-layer GNNs and is transferable to other GNNs.

\vspace{+0.5mm}
\noindent{\bf Acknowledgments.} 
We thank the anonymous reviewers for their constructive feedback. 
Wang is supported by the Cisco Research Award, National Science
Foundation under grant Nos. NSF-2216926 and NSF-2241713. Pang is supported in part by the Natural Science Foundation of Jiangxi Province of China (20232BAB212025) and High-level and Urgently Needed Overseas Talent Programs of Jiangxi Province (20232BCJ25024).

\noindent {\bf Ethical considerations:} Our work studies the vulnerability of graph neural networks, and it could probably have both a negative and a positive impact. 
From the negative side, our findings will inspire attackers to perform malicious activities against real-world systems. For instance, a malicious user in a social network (e.g, Twitter) can leverage our attack to make him avoid being detected by the malicious user detection system. Then, he can perform malicious activities, e.g., spreading fake news and distributing 
phishing attacks among the social network. 
From the positive side, our work will inspire 
following works to design more robust graph neural networks against adversarial attacks. 
All datasets and codes we used in the paper are publicly available. Our work is mainly for research purpose and complies with ethical standards. Therefore, it does not have any negative ethical impact on society.

%\balance
\bibliographystyle{ACM-Reference-Format}
\bibliography{refs,refs_certified}

% \section{Ethical Implications}

\appendix
\onecolumn
\clearpage

\begin{algorithm}[t]
\caption{Efficient calculation of label influence via DFS}
\begin{flushleft}
{\bf Input:}  
{Path length $K$+1, the node $n$ on the path, cumulative sum of weights $s$, target node $v$ and label $y_v$, target label $c$.} \\
{\bf Output:} {
Difference between the label influence of label-$y_v$ nodes and label-$c$ nodes.}
\end{flushleft}

\begin{algorithmic}[1]
\State $I_{y_v} \leftarrow 0, \ I_c \leftarrow 0$ \ 
\Function{LabelInfluence}{K, p, s}
    \If{$K=0$}
        \State {$I_{y_v} \leftarrow I_{y_v}+ s \cdot y_{v, y_v}$};
        {$I_c \leftarrow I_c + s \cdot y_{v, c}$};
        \Return
    \EndIf
    \For{$u \in \Gamma_p$}
        \State {$w = d_{p}^{-\frac{1}{2}}d_{u}^{-\frac{1}{2}}$};
        {$LabelInfluence(K-1, u, s \cdot w)$};
    \EndFor
\EndFunction

\State \Return $I_c - I_{y_v}$

\end{algorithmic}
\label{alg:cal_LI}
\end{algorithm}

\begin{algorithm}[!t]
\caption{(Label) influence-based targeted evasion attack}
\begin{flushleft}
{\bf Input:}  
{Adj. mat. $\mathbf{A}$, layer $K$, 
target node $v$, target label $c$, 
budget $\Delta$.} \\
{\bf Output:} $success$ (I.e., {$v$ attacks successfully or not})
\end{flushleft}

\begin{algorithmic}[1]
\State $AE \leftarrow 0$, $success \leftarrow False$, $\tilde{\mathbf{A}} \leftarrow \mathbf{A}$;
\State $\mathcal{N}_A, \mathcal{N}_B \leftarrow$ Find two candidate node sets from $\mathbf{A}$. 
\For{$a \in \mathcal{N}_A$}
    {$\Delta I_A(a) = d_{v}^{-\frac{1}{2}}d_{a}^{-\frac{1}{2}} \cdot LabelInfluence(K-1, a, 1)$}
\EndFor
\For{$b \in \mathcal{N}_B$}
    {$\Delta I_B(b) = d_{v}^{-\frac{1}{2}}d_{b}^{-\frac{1}{2}} \cdot LabelInfluence(K-1, b, 1)$}
\EndFor
\While{AE $< \Delta$}
    \State$C_A \leftarrow LabelInfluence(K, v, 1)$; // $d_v \leftarrow  d_v + 1$
    \State$C_B \leftarrow LabelInfluence(K, v, 1)$; // $d_v \leftarrow  d_v - 1$
    
    \State $u^* \leftarrow \textrm{argsort}(\{C_A+\Delta I_A\} \cup  \{C_B - \Delta I_B)\}$; 
    \If{$u^* \in \mathcal{N}_A$ }
        % \State 
        {$\tilde{\mathbf{A}} \leftarrow \mathbf{A} + (v,u^*)$; }
    \Else %if{$u^* \in \mathcal{N}_B$ }  
        % \State 
        { $\tilde{\mathbf{A}} \leftarrow \mathbf{A} - (v,u^*)$;}
    \EndIf
    \State $AE \leftarrow AE+1$
\EndWhile

\If{($f_{\Theta^*}(\tilde{\mathbf{A}})_{v,c}-f_{\Theta^*}(\tilde{\mathbf{A}})_{v,y_v})>0$}
    {
        $success \leftarrow True$; 
    }
\EndIf

\State \Return $success$
\end{algorithmic}
\label{alg:our_attack}
\end{algorithm}

\subsection*{Proof of Theorem~\ref{thm:relate}}
\label{sec:proof_themorem1}

We mainly focus on GCN, as SGC is a special case of GCN and the proof also applies. 
Our proof is based on the Lemma~\ref{lemma:gcn_influence} in Section~\ref{sec:relation}. 
Based on Equation (\ref{derivate_h_v_k}) in Lemma~\ref{lemma:gcn_influence}, the feature-label influence in Equation (\ref{feature_influence}) can be expressed as follows:
{
\begin{equation}
\begin{aligned}
    I_{fl}(v,u;K) 
    &= {\mathbf{1}_{y_u}^T [\frac{\partial \mathbf{h}_v^{(K)}}{\partial \mathbf{h}_u^{(0)}}] \mathbf{h}_u^{(0)}} 
    = {\mathbf{1}_{y_u}^T [{\prod_{l=K}^{1} \mathbf{W}^{(l)}} \cdot \rho \cdot  \sum_{p=1}^{\varPsi_{v \rightarrow u}}\prod_{l=K}^1{a_{v_p^{l},v_p^{l-1}}}] \mathbf{h}_u^{(0)}} \\
    &= \rho \cdot {\mathbf{1}_{y_u}^T [{\prod_{l=K}^{1} \mathbf{W}^{(l)}}] \mathbf{h}_u^{(0)} \cdot  \sum_{p=1}^{\varPsi_{v \rightarrow u}}\prod_{l=K}^1{a_{v_p^{l},v_p^{l-1}}}} 
    = C \cdot \sum_{p=1}^{\varPsi_{v \rightarrow u}}\prod_{l=K}^1{a_{v_p^{l},v_p^{l-1}}}, \\
\end{aligned}
\end{equation}
}%
where $C= \rho \mathbf{1}_{y_u}^T [{\prod_{l=1}^{K} \mathbf{W}^{(l)}}] \mathbf{h}_u^{(0)} $ is a constant for a given GCN. 

Comparing GCN with LP, we can find 
their iteration processes are similar, except that LP has no model parameters (which is constant for a trained GCN model). 
Specifically, we can calculate the label influence $I_l$ as follows: 
{
\begin{equation}
\begin{aligned}
    I_l(v,u;K) 
    = \frac{\partial y_v^{(K)}}{\partial y_u^{(0)}}
    = \sum_{p=1}^{\varPsi_{v \rightarrow u}}{\prod_{l=K}^1{a_{v_p^{l},v_p^{l-1}}}}.
\end{aligned}
\end{equation}
}

Hence, we can build the following relationship between the feature-label influence in GNN and label influence in LP: 
\begin{equation}
\label{influence_relationship}
    I_{fl}(v,u;K) = C \cdot I_l(v,u;K).    
\end{equation}

\subsection*{Proof of Theorem~\ref{thm:attack_label_inf}}
\label{sec:proof_thm2}

Let $\tilde{\mathbf{h}}_v^{(K)}$ be $u$'s final node representation after the attack. 
The attack's objective function in Equation (\ref{eqn:attack}) is equivalent to the following form:
{
\begin{align*}
    & \max\nolimits_{\tilde{\mathbf{A}}_v} \big([\sum\nolimits_{u\in \tilde{\mathbf{\Lambda}}_v^{(K)}}\frac{\partial \tilde{\mathbf{h}}_v^{(K)}}{\partial \mathbf{h}_u^{(0)}} \cdot \mathbf{h}_u^{(0)}]_c - [\sum\nolimits_{u\in \tilde{\mathbf{\Lambda}}_v^{(K)}}\frac{\partial \tilde{\mathbf{h}}_v^{(K)}}{\partial \mathbf{h}_u^{(0)}} \cdot \mathbf{h}_u^{(0)}]_{y_v}\big) % \notag \\
    \quad \, s.t., \quad \sum\nolimits_{s} |\tilde{A}_{v,s} - A_{v,s}| \leq \Delta.
\end{align*}
}%

Based on 
Assumption
~\ref{assump:fea_indep}, 
we further 
deduce attack's objective function in Equation (\ref{obj_h_der}) 
as follows:
{
\small
\begin{equation}
\begin{aligned}
\label{hyp1_obj}
    & [\sum_{u\in \tilde{\mathbf{\Lambda}}_v^{(K)}}\frac{\partial \tilde{\mathbf{h}}_v^{(K)}}{\partial \mathbf{h}_u^{(0)}} \cdot \mathbf{h}_u^{(0)}]_c - [\sum_{u\in \tilde{\mathbf{\Lambda}}_v^{(K)}}\frac{\partial \tilde{\mathbf{h}}_v^{(K)}}{\partial \mathbf{h}_u^{(0)}} \cdot \mathbf{h}_u^{(0)}]_{y_v} \notag \\
    & \approx [\sum_{\substack{u\in \tilde{\mathbf{\Lambda}}_v^{(K)}, y_u=c}}\frac{\partial \tilde{\mathbf{h}}_v^{(K)}}{\partial \mathbf{h}_u^{(0)}} \cdot \mathbf{h}_u^{(0)} ]_c - [\sum_{\substack{z\in \tilde{\mathbf{\Lambda}}_v^{(K)}, y_z=y_v}}\frac{\partial \tilde{\mathbf{h}}_v^{(K)}}{\partial \mathbf{h}_z^{(0)}} \cdot \mathbf{h}_z^{(0)}]_{y_v} \\
    & = \sum_{\substack{u\in \tilde{\mathbf{\Lambda}}_v^{(K)}, y_u=c}} [\frac{\partial \tilde{\mathbf{h}}_v^{(K)}}{\partial \mathbf{h}_u^{(0)}} \cdot \mathbf{h}_u^{(0)} ]_c - \sum_{\substack{z\in \tilde{\mathbf{\Lambda}}_v^{(K)}, y_z=y_v}} [\frac{\partial \tilde{\mathbf{h}}_v^{(K)}}{\partial \mathbf{h}_z^{(0)}} \cdot \mathbf{h}_z^{(0)} ]_{y_v} \\
    & = \sum_{\substack{u\in \tilde{\mathbf{\Lambda}}_v^{(K)}, y_u=c}} [\mathbf{1}_{y_u}^T \cdot \frac{\partial \tilde{\mathbf{h}}_v^{(K)}}{\partial \mathbf{h}_u^{(0)}} \cdot \mathbf{h}_u^{(0)} ] - \sum_{\substack{z\in \tilde{\mathbf{\Lambda}}_v^{(K)}, y_z=y_v}} [\mathbf{1}_{y_z}^T \frac{\partial \tilde{\mathbf{h}}_v^{(K)}}{\partial \mathbf{h}_z^{(0)}} \cdot \mathbf{h}_z^{(0)}].
\end{aligned}
\end{equation}
}%

\noindent Note that the two terms $[\mathbf{1}_{y_u}^T \cdot \frac{\partial \tilde{\mathbf{h}}_v^{(K)}}{\partial \mathbf{h}_u^{(0)}} \cdot \mathbf{h}_u^{(0)} ]$ and $[\mathbf{1}_{y_z}^T \cdot \frac{\partial \tilde{\mathbf{h}}_v^{(K)}}{\partial \mathbf{h}_z^{(0)}} \cdot \mathbf{h}_z^{(0)}]$ are exactly the feature-label influence after the attack, %attacking the graph. 
and we denote them as $\tilde{I}_{fl}(v,u;K)$ and $\tilde{I}_{fl}(v,z;K)$, respectively.
Based on the relationship between the feature-label influence and label influence in Equation (\ref{influence_relationship}),
we thus have the following attack's objective function in terms of label influence:
{
\begin{equation*}
\begin{aligned}
\label{hyp1}
    & \max_{\tilde{\mathbf{A}}_v} \big(\sum_{\substack{u\in \tilde{\mathbf{\Lambda}}_v^{(K)}, y_u=c}}\tilde{I}_{fl}(v,u;K)
    -\sum_{\substack{z\in \tilde{\mathbf{\Lambda}}_v^{(K)}, y_z=y_v}} \tilde{I}_{fl}(v,z;K) \big) % \\ %\cdot h_z) \\
    \Leftrightarrow & \max_{\tilde{\mathbf{A}}_v} \big (\sum_{\substack{u\in \tilde{\mathbf{\Lambda}}_v^{(K)}, y_u=c}}\tilde{I}_{l}(v,u;K) %\cdot \mathbf{h}_u 
    -\sum_{\substack{z\in \tilde{\mathbf{\Lambda}}_v^{(K)} , y_z=y_v}}\tilde{I}_{l}(v,z;K) \big). \\
\end{aligned}
\end{equation*}
}%

\subsection*{Efficient Label Influence Calculation}

First, we have the following two observations:
\begin{itemize}[leftmargin=*]
\item When adding an edge between $v$ and 
$a \in \mathcal{N}_A$, 
we will have 
new paths $\{v, a, \cdots, u \}$ from $v$ to $u$ passing through $a$. 
We denote the nodes in the new paths 
within $v$'s $K$-hop neighbors as $\Delta \mathbf{\Lambda}_{v,a}^{(K)}$, and  
note that these nodes are within $a$'s $(K-1)$-hop neighbors in the clean graph, i.e., $\Delta \mathbf{\Lambda}_{v,a}^{(K)} \subset \mathbf{\Lambda}_a^{(K-1)}$.
Moreover, $\tilde{\mathbf{\Lambda}}_v^{(K)} = \mathbf{\Lambda}_v^{(K)} \cup {\mathbf{\Lambda}}_a^{(K-1)}$. 

\item When deleting an edge between $v$ and $b \in \mathcal{N}_B$, we will 
remove existing paths $\{v, b, \cdots, u \}$ from $v$ to $u$  
passing through $b$. 
% For notation simplicity, we also 
We denote the deleted nodes within $v$'s $K$-hop neighbors as $\Delta \mathbf{\Lambda}_{v,b}^{(K)}$, and  
these nodes are within $b$'s $(K-1)$-hop neighbors in the clean graph, i.e., $\Delta \mathbf{\Lambda}_{v,b}^{(K)} \subset \mathbf{\Lambda}_b^{(K-1)}$. 
Moreover, $\tilde{\mathbf{\Lambda}}_v^{(K)} = \mathbf{\Lambda}_v^{(K)} \setminus {\mathbf{\Lambda}}_b^{(K-1)}$. 
\end{itemize}

\noindent Based on the 
two 
observations, we can 
split each label influence 
in  Equation (\ref{obj:attack_label_inf}) into two parts: an \emph{approximate constant label influence} defined on the clean graph and an \emph{approximate label influence} defined on the $(K-1)$-hop neighbors for each node in $\mathcal{N}_A \cup \mathcal{N}_B$. 

We first consider {\bf adding ({\color{red} +})} an edge between node $u$ and node $a \in \mathcal{N}_A$. Specifically, we have: 
{
\small
\begin{align*}
    & \sum\nolimits_{\substack{u\in \tilde{\mathbf{\Lambda}}_v^{(K)}, y_u=c}}\tilde{I}_{l}(v,u;K)%\cdot \mathbf{h}_u
    -\sum\nolimits_{\substack{z\in \tilde{\mathbf{\Lambda}}_v^{(K)}, y_z=y_v}}\tilde{I}_{l}(v,z;K) \\
    = & [ \sum_{\substack{u\in {{\mathbf{\Lambda}}_v^{(K)}}, y_u=c}}\tilde{I}_{l}(v,u;K) \, {\color{red}+}
    \sum_{\substack{u\in \Delta {\mathbf{\Lambda}}_{v,a}^{(K)},  y_u=c}}\tilde{I}_{l}(v,u;K) ]  - [ \sum_{\substack{z\in {{\mathbf{\Lambda}}_v^{(K)}}, y_z=y_v}}\tilde{I}_{l}(v,z;K) \, {\color{red}+} \sum_{\substack{z\in \Delta {\mathbf{\Lambda}}_{v,a}^{(K)}, y_z=y_v}}\tilde{I}_{l}(v,z;K) ]    \\
    =& [ \sum_{\substack{u\in {{\mathbf{\Lambda}}_v^{(K)}}, y_u=c}}\tilde{I}_{l}(v,u;K) - \sum_{\substack{z\in {{\mathbf{\Lambda}}_v^{(K)}}, y_z=y_v}}\tilde{I}_{l}(v,z;K) ] { + } [ \sum_{\substack{u\in \Delta {\mathbf{\Lambda}}_{v,a}^{(K)},  y_u=c}}\tilde{I}_{l}(v,u;K) - \sum_{\substack{z\in \Delta {\mathbf{\Lambda}}_{v,a}^{(K)}, y_z=y_v}}\tilde{I}_{l}(v,z;K) ]    \\
    =& [ \sum_{\substack{u\in {{\mathbf{\Lambda}}_v^{(K)}},  y_u=c}}\sum_{p=1}^{{\varPsi}_{v \rightarrow u}}
    \prod_{l=K}^{1}\tilde{d}^{-\frac{1}{2}}_{v_p^l} \tilde{d}^{-\frac{1}{2}}_{v_p^{l-1}} - \sum_{\substack{z\in {{\mathbf{\Lambda}}_v^{(K)}}, y_z=y_v}}\sum_{p=1}^{{\varPsi}_{v \rightarrow z}}
    \prod_{l=K}^{1}\tilde{d}^{-\frac{1}{2}}_{v_p^l} \tilde{d}^{-\frac{1}{2}}_{v_p^{l-1}}]  { + } [ \sum_{\substack{u\in \Delta{{\mathbf{\Lambda}}_{v,a}^{(K)}},  y_u=c}}\sum_{p=1}^{\tilde{\varPsi}_{v \rightarrow u}}
    \prod_{l=K}^{1}\tilde{d}^{-\frac{1}{2}}_{v_p^l} \tilde{d}^{-\frac{1}{2}}_{v_p^{l-1}}  - \sum_{\substack{z\in \Delta{{\mathbf{\Lambda}}_{v,a}^{(K)}},  y_z=y_v}}\sum_{p=1}^{\tilde{\varPsi}_{v \rightarrow z}}
    \prod_{l=K}^{1}\tilde{d}^{-\frac{1}{2}}_{v_p^l} \tilde{d}^{-\frac{1}{2}}_{v_p^{l-1}} ]  \\
    \approx & [ \sum_{\substack{u\in {{\mathbf{\Lambda}}_v^{(K)}},  y_u=c}}\sum_{p=1}^{{\varPsi}_{v \rightarrow u}}
    \prod_{l=K}^{1} {d}^{-\frac{1}{2}}_{v_p^l} {d}^{-\frac{1}{2}}_{v_p^{l-1}} - \sum_{\substack{z\in {{\mathbf{\Lambda}}_v^{(K)}},  y_z=y_v}}\sum_{p=1}^{{\varPsi}_{v \rightarrow z}}
    \prod_{l=K}^{1}{d}^{-\frac{1}{2}}_{v_p^l} {d}^{-\frac{1}{2}}_{v_p^{l-1}}] \\
    & + [ \sum_{\substack{u\in \Delta{{\mathbf{\Lambda}}_{v,a}^{(K)}},  y_u=c}}\sum_{p=1}^{\tilde{\varPsi}_{v \rightarrow u}}
    \prod_{l=K}^{1}{d}^{-\frac{1}{2}}_{v_p^l} {d}^{-\frac{1}{2}}_{v_p^{l-1}}  - \sum_{\substack{z\in \Delta{{\mathbf{\Lambda}}_{v,a}^{(K)}},  y_z=y_v}}\sum_{p=1}^{\tilde{\varPsi}_{v \rightarrow z}}
    \prod_{l=K}^{1}{d}^{-\frac{1}{2}}_{v_p^l} {d}^{-\frac{1}{2}}_{v_p^{l-1}} ]  \\
    =&  C_A  + \Delta {{I}_{A}}(a),
\end{align*}
}%
where 
\begin{itemize}[leftmargin=*]
\item 
$\sum_{\substack{u\in \Delta {\mathbf{\Lambda}}_{v,a}^{(K)}, y_u=c}}\tilde{I}_{l}(v,u;K)$, $\sum_{\substack{z\in \Delta {\mathbf{\Lambda}}_{v,a}^{(K)},  y_z=y_v}}\tilde{I}_{l}(v,z;K)$ represent the label influence of the label-$c$ nodes and label-$y_v$ nodes within the $(K-1)$-hop neighbors of $a$ after adding the edge $(v, a)$, respectively. 
Note that the degree of most nodes in these paths do not change, except that the degree of the target node $v$ and the node $a$ % certain nodes in these paths 
increases by 1 due to the added edge. Here, for efficient computation, we assume all nodes in ${{\mathbf{\Lambda}}_v^{(K)}}$, except the target node $v$, do not change the node degree. That means,  we approximately set 
$\tilde{d}_{v_p^l} = {d}_{v_p^l}$ and $\tilde{d}_{v_p^{l-1}} = {d}_{v_p^{l-1}}$, for $v_p^{l} \neq v$ and $v_p^{l-1} \neq v$. 

\item $C_A$ is the approximate constant label influence defined on the \emph{clean graph} and only needs to be calculated once.

\item $\Delta I_{A}(a)$ is the approximate label influence caused by 
the perturbed edge $(v, a)$, $\forall a \in \mathcal{N}_A$. We can calculate $\Delta I_{A}(a)$ for each attack edge $(v, a)$ in advance. 
\end{itemize}
When {\bf removing} an edge between nodes $u$ and  $b \in \mathcal{N}_B$, we replace ${\color{red}+}$ with ${\color{red}-}$ and assume all nodes' degree do not change, except the target node's degree decreases by 1.
We denote the approximate constant label influence on the clean graph as $C_B$ and the label influence per perturbed edge $(v, b)$, $\forall b \in \mathcal{N}_B$, as  $\Delta I_B(b)$, respectively.
Similarly, $C_B$ only needs to be calculated once and $\Delta I_{B}(b)$ can be calculated  for each attack edge $(v, b)$
in advance.

\begin{figure}[!t]
\centering
\includegraphics[width=0.5\linewidth]{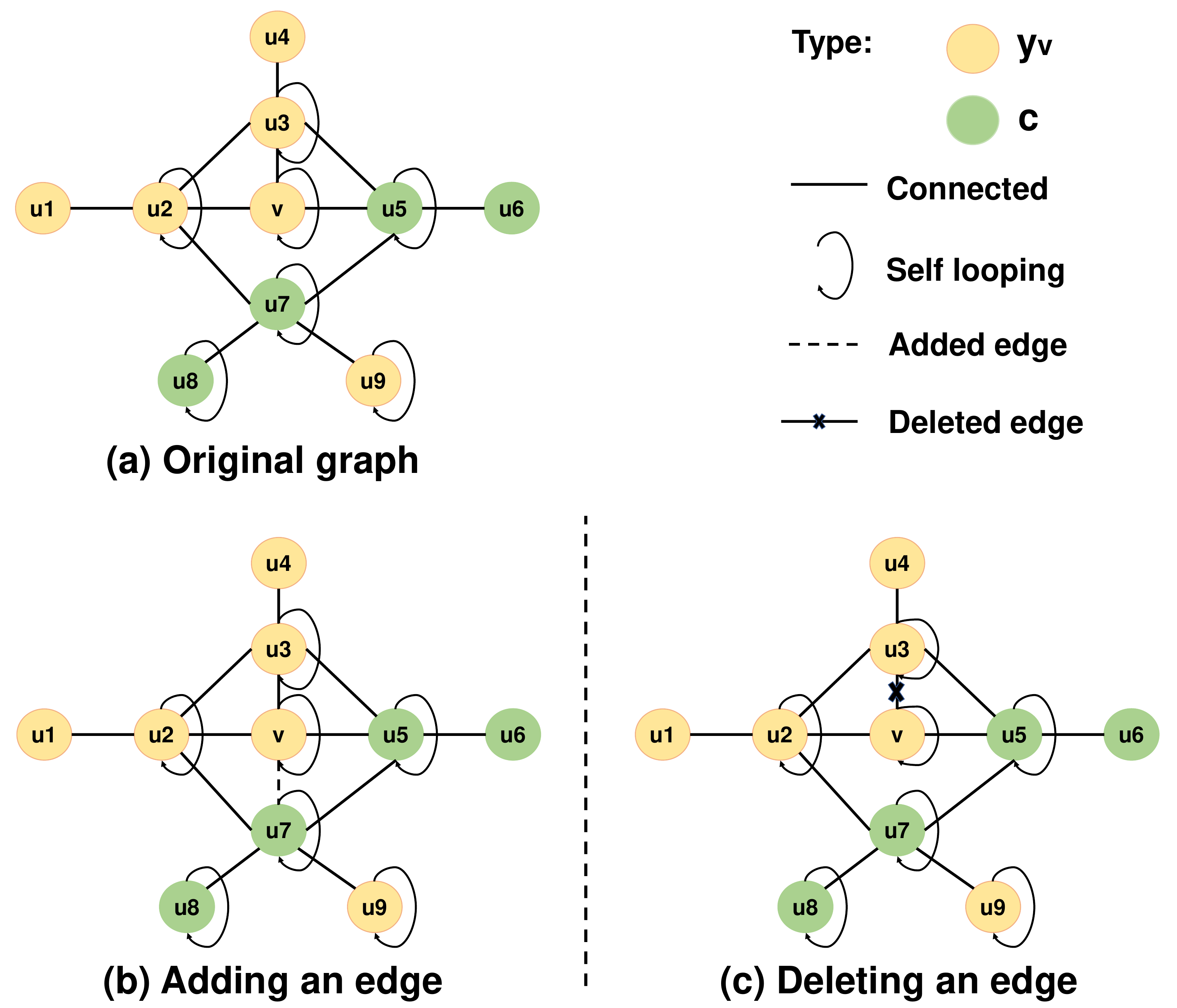}
\caption{An example for calculating label influence.} 
\label{toy_example}
\vspace{-2mm}
\end{figure}

\section*{Example of calculating influence 
}
Now we show 
an example for exact and approximate label influence calculation.
We take Figure \ref{toy_example} as an example
to illustrate how label influence terms are calculated when adding a new edge or deleting an existing edge when $K=2$. 
The original graph is shown in Figure \ref{toy_example}(a). 
There are two classes of nodes, i.e., green color and yellow color. 
$v$ is the target node with a label-$y_v$.
The candidate nodes for $v$ to add edges are $\mathcal{N}_A = \{u_6, u_7, u_8 \}$,
and 
the candidate nodes for $v$ to delete edges are $\mathcal{N}_B = \{u_2, u_3\}$ with label yellow. 

In the clean graph, we have $\mathbf{\Lambda}_v^{(2)} = [v, u_1, u_2, u_3, u_4, u_5, u_6, u_7] $; 
$\{ u\in {\mathbf{\Lambda}}_v^{(2)}, y_u=c \} = \left \{u_5,u_6,u_7 \right \}$; $ \{ z\in {\mathbf{\Lambda}}_v^{(2)}, y_z=y_v \} =  \left \{ u_1, u_2,u_3, u_4 \right \}$.

Then, we have all the paths 
\begin{itemize}[leftmargin=*]
\item {\color{red} (a)} $\{ \Psi_{v \rightarrow u} |  u \in  {\mathbf{\Lambda}}_v^{(2)}, y_u=c  \} = \{ v-u_5-u_5, v-v-u_5,  v-u_3-u_5, v-u_5-u_6, v-u_2-u_7, v-u_5-u_7\}$; 

\item {\color{red} (b)} $\{ \Psi_{v \rightarrow z} | z\in {\mathbf{\Lambda}}_v^{(2)}, y_z=y_v \} = \{v-v-v, v-u_5-v, v-u_2-v, v-u_3-v, v-u_2-u_1, v-u_2-u_2, v-v-u_2, v-u_3-u_2, v-u_3-u_3, v-v-u_3, v-u_2-u_3, v-u_5-u_3, v-u_3-u_4\}$
\end{itemize}

\subsubsection*{Adding an edge} 
We first compute the exact label influence and then calculate the approximate label influence.

Figure \ref{toy_example}(b) is the graph structure after adding an edge between $u_7$ and $v$. 
In the perturbed graph, we have:
$\tilde{\mathbf{\Lambda}}_v^{(2)} = [v, u_1, u_2, u_3, u_4, u_5, u_6, \\ u_7, u_8, u_9] $;  
$\Delta \mathbf{\Lambda}_{v,u_7}^{(2)} = \{v, u_2, u_5, u_7, u_8, u_9 \}$; 
$\{ u\in \Delta {\mathbf{\Lambda}}_{v,u_7}^{(2)}, y_u=c \} = \left \{u_5, u_7, u_8 \right \}$; 
$\{ z\in \Delta {\mathbf{\Lambda}}_{v,u_7}^{(2)}, y_z=y_v \} =  \left \{ v, u_2, u_9 \right \}$. 

 The new added paths are passing through $u_7$. Specifically, all the new added paths are:
\begin{itemize}[leftmargin=*]
\item {\color{red} (c)}
$\{ \varPsi'_{v \rightarrow u} | u\in \Delta {\mathbf{\Lambda}}_{v,u_7}^{(2)}, y_u=c \} = \{v-v-u_7, v-u_7-u_5, v-u_7-u_7, v-u_7-u_8\}$;

\item {\color{red} (d)} $\{ \varPsi'_{v \rightarrow z} | z\in \Delta {\mathbf{\Lambda}}_{v,u_7}^{(2)}, y_z=y_v \} = \{v-u_7-v, v-u_7-u_2, v-u_7-u_9\}$.
\end{itemize}

\noindent {\bf Exact label influence calculation.} 
We first precisely calculate the label influence as follows:
{
\begin{align*}
\label{adding_I_f}
    & \sum\nolimits_{\substack{u\in \tilde{\mathbf{\Lambda}}_v^{(2)},  y_u=c}}\tilde{I}_l(v,u;2)%\cdot h_u
    -\sum\nolimits_{\substack{z\in \tilde{\mathbf{\Lambda}}_v^{(2)},  y_z=y_v}}\tilde{I}_l(v,z;2) \\
    = & \llbracket ({d}_{u_5}^{-\frac{3}{2}}{d_v'^{-\frac{1}{2}}} + {d}_{u_5}^{-\frac{1}{2}}{d_v'^{-\frac{3}{2}}} + {d}_{u_5}^{-\frac{1}{2}}{d}_{u_3}^{-1}{d_v'^{-\frac{1}{2}}} + {d}_{u_6}^{-\frac{1}{2}}{d}_{u_5}^{-1}{d_v'^{-\frac{1}{2}}} + {d_{u_7}'^{-\frac{1}{2}}}{d}_{u_2}^{-1}{d_v'^{-\frac{1}{2}}} + {d_{u_7}'^{-\frac{1}{2}}}{d}_{u_5}^{-1}{d_v'^{-\frac{1}{2}}})  \textrm{ \color{red} (a)} \\
    & - ({d_v'^{-2}} + {d_v'^{-1}}{d}_{u_5}^{-1} +  {d_v'^{-1}}{d}_{u_2}^{-1} + {d_v'^{-1}}{d}_{u_3}^{-1} + {d}_{u_1}^{-\frac{1}{2}}{d}_{u_2}^{-1}{d_v'^{-\frac{1}{2}}}  + {d}_{u_2}^{-\frac{3}{2}}{d_v'^{-\frac{1}{2}}} + {d}_{u_2}^{-\frac{1}{2}}{d_v'^{-\frac{3}{2}}} + {d}_{u_2}^{-\frac{1}{2}}{d}_{u_3}^{-1}{d_v'^{-\frac{1}{2}}} 
    + {d}_{u_3}^{-\frac{3}{2}}{d_v'^{-\frac{1}{2}}} + {d}_{u_3}^{-\frac{1}{2}}{d_v'^{-\frac{3}{2}}} + {d}_{u_3}^{-\frac{1}{2}}{d}_{u_2}^{-1}{d_v'^{-\frac{1}{2}}} 
    \\
    & \quad  
    + {d}_{u_3}^{-\frac{1}{2}}{d}_{u_5}^{-1}{d_v'^{-\frac{1}{2}}} +
    {d}_{u_4}^{-\frac{1}{2}}{d}_{u_3}^{-1}{d_v'^{-\frac{1}{2}}}) \rrbracket \textrm{{ \color{red} (b)}}\\
    &+ \llbracket ({d_{u_7}'^{-\frac{1}{2}}}{d_v'^{-\frac{3}{2}}} + {d}_{u_5}^{-\frac{1}{2}}{d_{u_7}'^{-1}}{d_v'^{-\frac{1}{2}}} + {d_{u_7}'^{-\frac{3}{2}}}{d_v'^{-\frac{1}{2}}} + {d}_{u_8}^{-\frac{1}{2}}{d_{u_7}'^{-1}}{d_v'^{-\frac{1}{2}}}) \textrm{{ \color{red} (c)}} \\
    & \quad - 
    ({d_v'^{-1}}{d_{u_7}'^{-1}} + {d}_{u_2}^{-\frac{1}{2}}{d_{u_7}'^{-1}}{d_v'^{-\frac{1}{2}}} + {d}_{u_9}^{-\frac{1}{2}}{d_{u_7}'^{-1}}{d_v'^{-\frac{1}{2}}} ) \rrbracket \textrm{{ \color{red} (d)}} \\
    =& C_A (exact) + I_A(u_7) \\
    =& \left(0.2563 - 0.5665\right) + \left(0.1530 - 0.1194\right) \\ 
    =& -0.3102 + 0.0336 = -0.2766.
\end{align*}
}%
Note that $C_A (exact) = -0.3102$ is the exact constant label influence in the original paths, and $I_A(u_7) = 0.0336$ is the exact label influence in the new added paths.

\noindent \emph{\bf Approximate label influence calculation.} 
When calculating the label influence in the original paths in 
the above Equation 
after adding the edge between $v$ and $u_7$, we notice that only paths including $u_7$ are affected, as $u_7$'s degree increases by 1. Here, for efficient calculation, we ignore the effect caused by $u_7$'s degree, i.e., we set $d_{u_7}'=d_{u_7}$, which means that we do not need to normalize the weights of edges associated with $u_7$. For the target node $v$, its degree increases by 1 and we still use  ${d}'_{v}={d}_{v}+1$. 

With this setting, we calculate 
the approximate constant label influence of the first term $\sum_{\substack{u\in \tilde{\mathbf{\Lambda}}_v^{(2)},  y_u=c}}\tilde{I}_l(v,u;2)$ is $C_A (approx) = -0.3032$, which is close to the precise value $-0.3102$. 
Moreover, we compute the approximate label influence for the second term $\sum_{\substack{z\in \tilde{\mathbf{\Lambda}}_v^{(2)},  y_z=y_v}}\tilde{I}_l(v,z;2)$, whose value is $I_A(approx, u_7) = 0.0299$ and 
close to the precise value 0.0336. 
Both the results 
verify the effectiveness of our efficient calculation of label influence  
for adding edges.

\subsubsection*{Deleting an edge} 
Figure \ref{toy_example}(c) is the graph after deleting an edge between $u_3$ and $v$. 
In the perturbed graph, we have 
$\tilde{\mathbf{\Lambda}}_v^{(2)} = [v, u_1, u_2, u_3, u_4, \\ u_5, u_6, u_7] $;
$\Delta \mathbf{\Lambda}_{v,u_3}^{(2)} = \{v, u_2, u_3, u_4, u_5 \}$; 
$\{ u\in \Delta {\mathbf{\Lambda}}_{v,u_3}^{(2)}, y_u=c \} = \left \{u_5 \right \}$; 
$\{ z\in \Delta {\mathbf{\Lambda}}_{v,u_3}^{(2)}, y_z=y_v \} =  \left \{ v, u_2, u_3, u_4 \right \}$. 
The new deleted paths are passing through $u_3$. Specifically, all the new deleted paths are:
    $\{ \varPsi'_{v \rightarrow u} | u\in \Delta {\mathbf{\Lambda}}_{v,u_3}^{(2)}, y_u=c \} = \{v-u_3-u_5\}$. 
    $\{ \varPsi'_{v \rightarrow z} | z\in \Delta {\mathbf{\Lambda}}_{v,u_3}^{(2)}, y_z=y_v\} = \{v-u_3-v, v-u_3-u_2, v-v-u_3, v-u_3-u_3, v-u_3-u_4\}$.

Similarly, when calculating the label influence in the original paths after removing the edge between $v$ and $u_3$, we notice that only paths including $u_3$ are affected, as $u_3$'s degree decreases by 1. Here, for efficient calculation, we ignore the effect caused by $u_3$'s degree, i.e., we set $d_{u_3}'=d_{u_3}$, which means that we do not need to normalize the weights of edges associated with $u_3$. For the target node $v$, its degree decreases by 1 and we still use  ${d}'_{v}={d}_{v}-1$.

With this setting, we calculate that the exact label influence of the first term is $C_B(exact) = -0.5423$ and its approximate value is $C_B(approx) = -0.5304$. Moreover, the exact label influence 
of the second term is $I_B(u_3) = -0.2860$, and its approximate label influence is $I_B(approx, u_3) = -0.2656$. The two approximate  values and precise values are also close.

\subsection*{More Experimental Results}
\label{app:results}

\begin{table}[!t]\renewcommand{\arraystretch}{1.0}
\centering
\caption{Dataset statistics.}
\centering
\begin{tabular}{|c|c|c|c|c|c|c|} 
\hline 
%\toprule 
{\bf Dataset}  & {\bf \#Nodes} & {\bf \#Edges} & {\bf \#Features} & {\bf \#Classes} \\ \hline
{\bf Cora}  &  {\bf 2,708} & {\bf 5,429} & {\bf 1,433}   & {\bf 7} \\ \hline
{\bf Citeseer}  &  {\bf 3,327} & {\bf 4,732} & {\bf 3,703}   & {\bf 6} \\ \hline
{\bf Pubmed}  &  {\bf 19,717} & {\bf 44,338} & {\bf 500}   & {\bf 3}  \\ \hline
{\bf OGB-arxiv}  &  {\bf 169,343} & {\bf 1,166,243} & {\bf 128}   & {\bf 40}  \\ \hline
\end{tabular}
%\vspace{-2mm}
\label{dataset_stat}
\end{table}

\begin{table}[t]%\renewcommand{\arraystretch}{1.1}
\centering
\caption{Fraction of label influences of target nodes' within $K$-hop neighbors that have the same label as the target node.}
\begin{tabular}{|c|c|c|c|c|c|c|c|c|c|}
\hline
{\bf Dataset} & \multicolumn{3}{c|}{\bf Cora} & \multicolumn{3}{c|}{\bf Citeseer} & \multicolumn{3}{c|}{\bf Pubmed} \\ \hline
 {\bf K}                 &  2     &    3   &  4     & 2      &   3    &   4    &  2     &  3     &   4    \\ \hline
     {\bf Two-layer GCN}             &   86\%    &   83\%    &  80\%     &  89\%    &  88\%     &   87\%    &  84\%     &   81\%    &   80\%    \\ \hline
     {\bf Two-layer SGC}             &   86\%    &   82\%    &  80\%     &  92\%    &  91\%     &   90\%    &  88\%     &   84\%    &   82\%    \\ \hline 
\end{tabular}
\label{tbl:infl_summarize}
% \vspace{-2mm}
\end{table}

\vspace{+0.5mm}  \noindent {\bf Verifying the correctness of Assumption~\ref{assump:fea_indep}}. 
In this experiment, we aim to validate Assumption~\ref{assump:fea_indep}. Specifically, in the three datasets, 
for each of the 100 target nodes, we summarize the label influences of all its $K$-hop neighbors and select $K=2,3,4$ in our experiments (The shortest paths in these datasets are  $\leq 4$). The results on two-layer GCN/SGC are shown in Table~\ref{tbl:infl_summarize}.  
We observe that, in all datasets, a majority, i.e., more than 80\%, of the total influences are from the nodes that have the same label as the target node. 
Such results validate   Assumption~\ref{assump:fea_indep} holds in general.

\begin{figure*}[!t]
\center
%\vspace{-2mm}
\subfigure[Cora]{\includegraphics[width=0.3\textwidth]{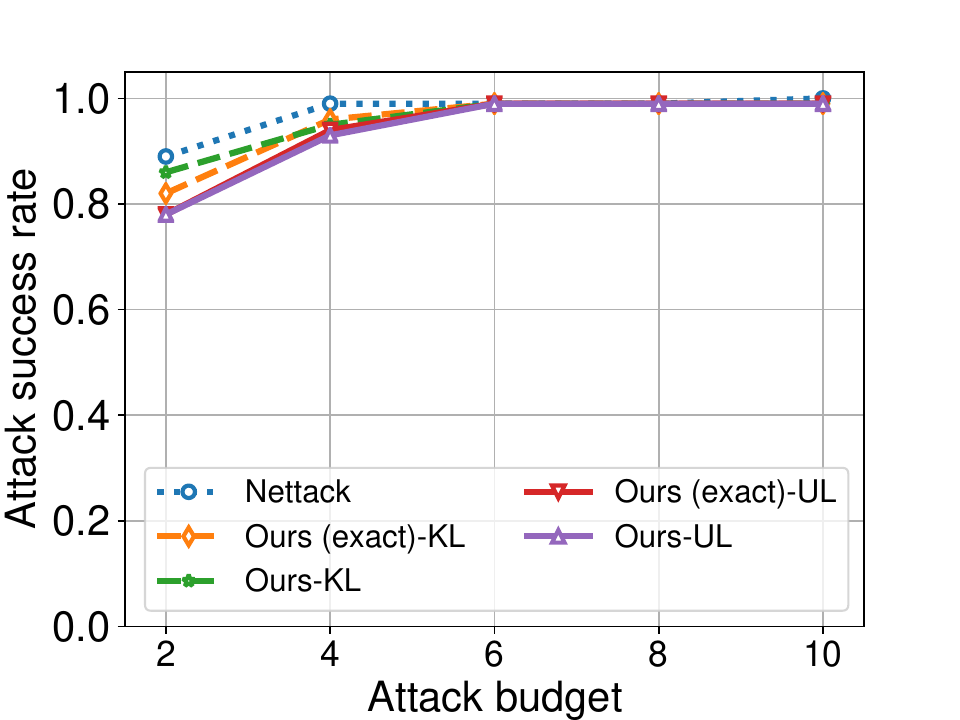}}
\subfigure[Citeseer]{\includegraphics[width=0.3\textwidth]{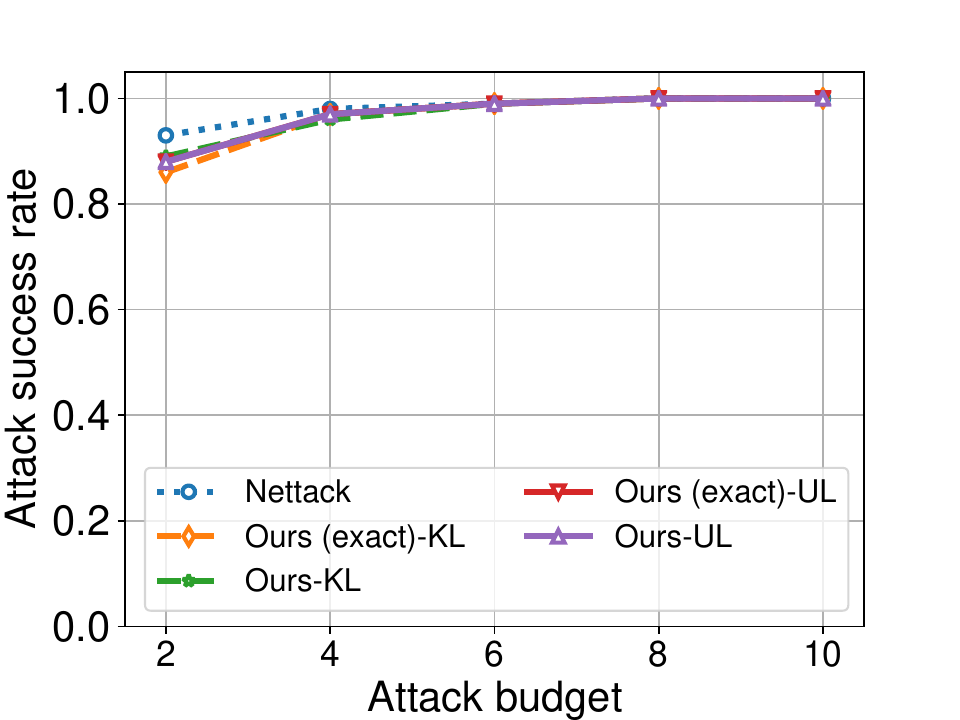}}
\subfigure[Pubmed]{\includegraphics[width=0.3\textwidth]{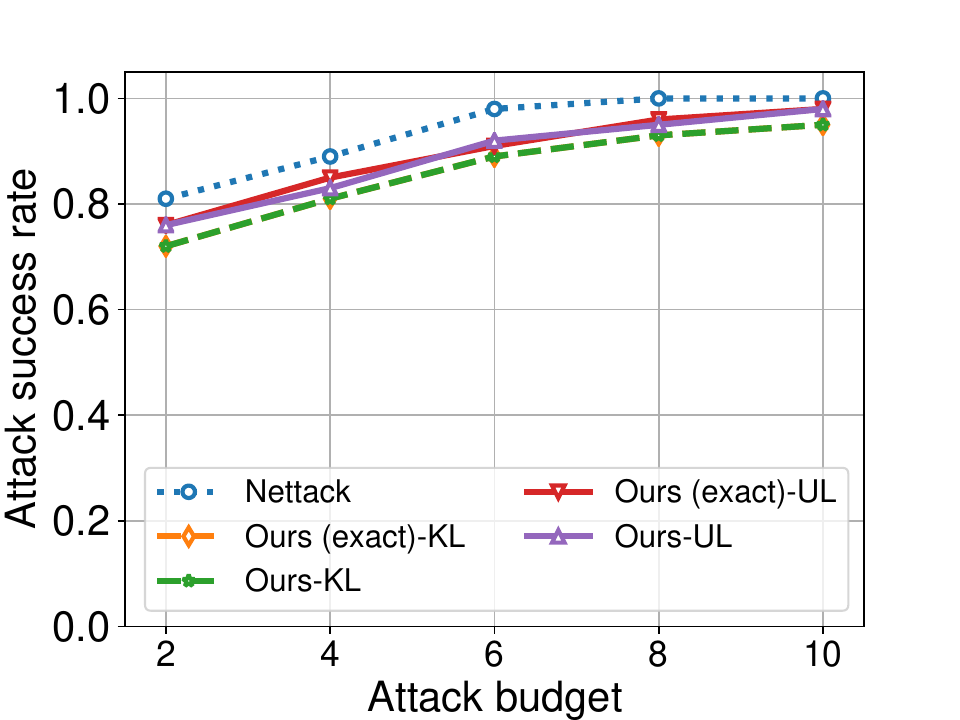}}
\vspace{-4mm}
\caption{Attack success rate vs. attack budget per target node on a two-layer SGC of all compared attacks on the three graphs.}
\label{fig:attack_2layer_SGC}
\vspace{-4mm}
\end{figure*}

\begin{figure*}[t]
\center
\subfigure[Cora]{\includegraphics[width=0.3\textwidth]{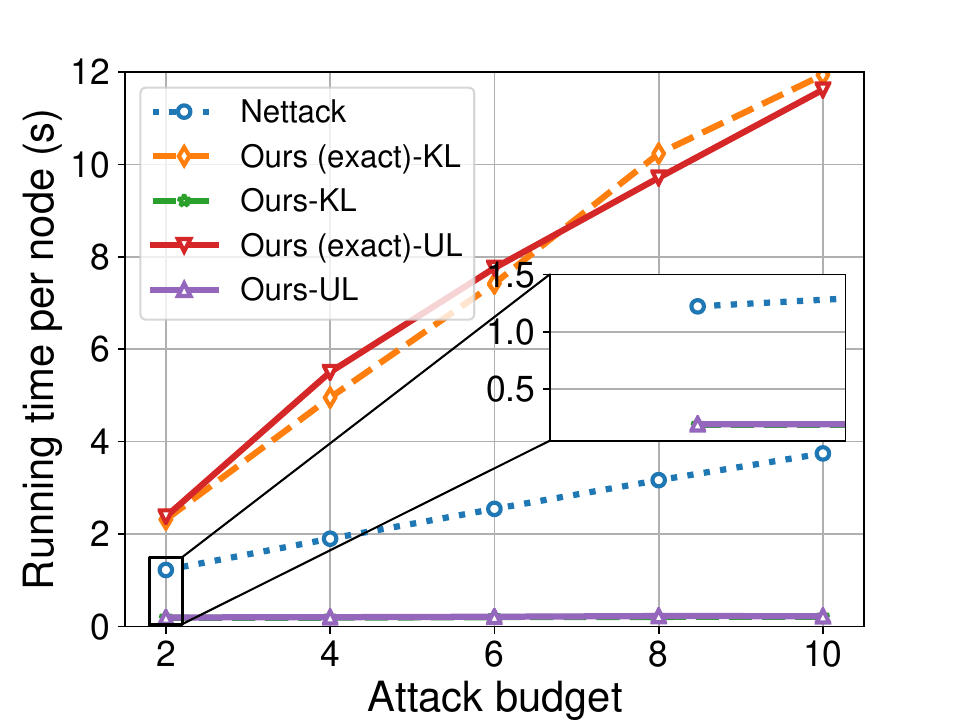}}
\subfigure[Citeseer]{\includegraphics[width=0.3\textwidth]{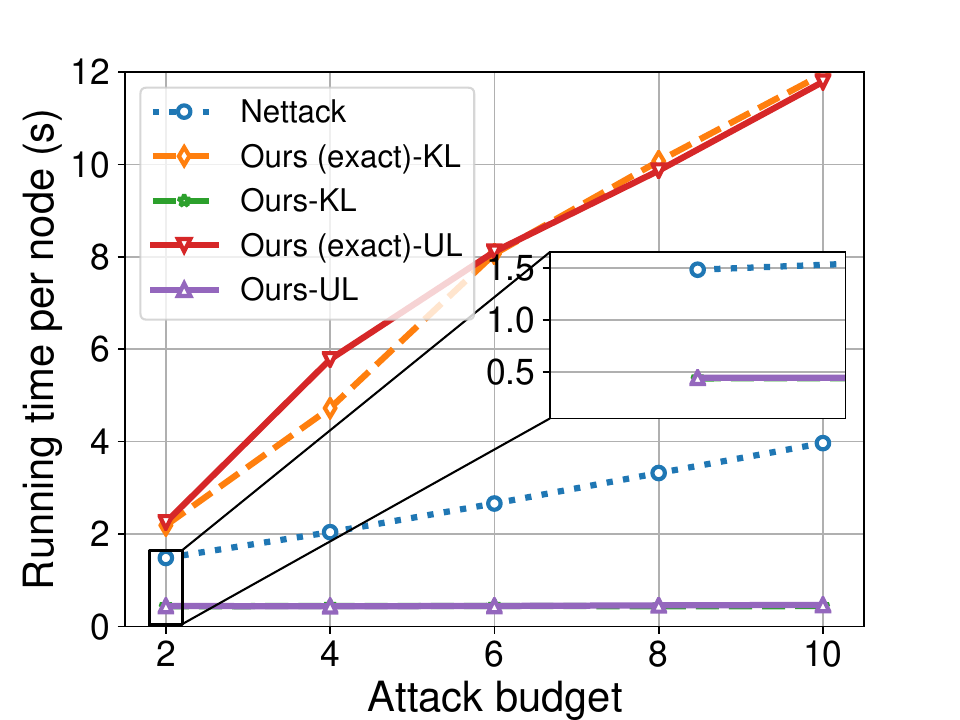}}
\subfigure[Pubmed]{\includegraphics[width=0.3\textwidth]{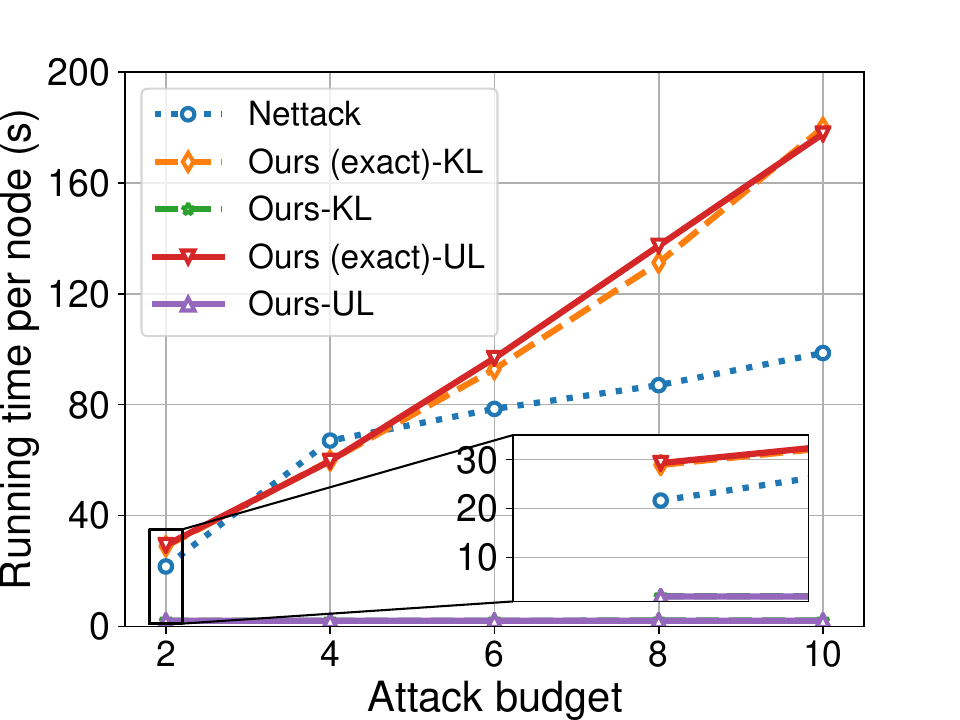}}
\vspace{-4mm}
\caption{Running time vs. attack budget per target node on two-layer SGC of all compared attacks on the three graphs.}
\label{fig:time_2layer_SGC}
\end{figure*}

\begin{figure}[!t]
  \centering
    \includegraphics[width=0.35\textwidth]{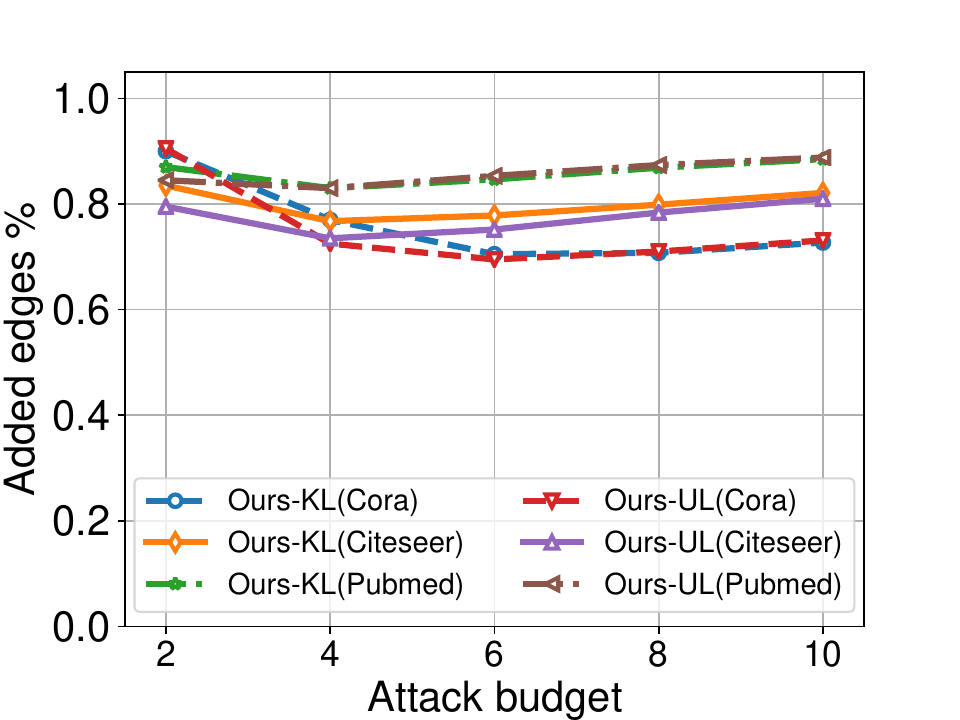} 
    \caption{Fraction of the added edges among all attack edges generated by our attacks against (a) two-layer GCN and (b) two-layer SGC vs. attack budget 
    per target node.}
    \label{fig:Added_edges_SGC}
\end{figure}

\begin{figure}[t]
\center
\subfigure[Attack success rate]{\includegraphics[width=0.35\textwidth]{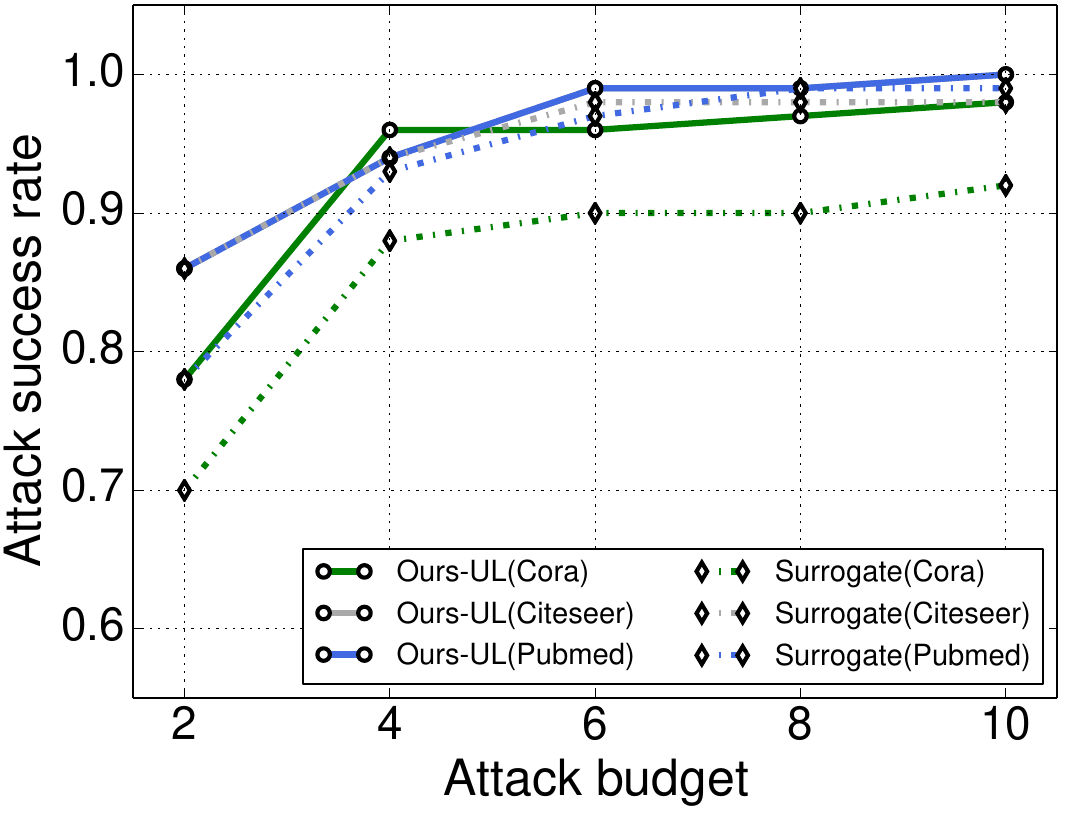} \label{fig:SGC_4layer_attack}}
\subfigure[Running time]{\includegraphics[width=0.35\textwidth]{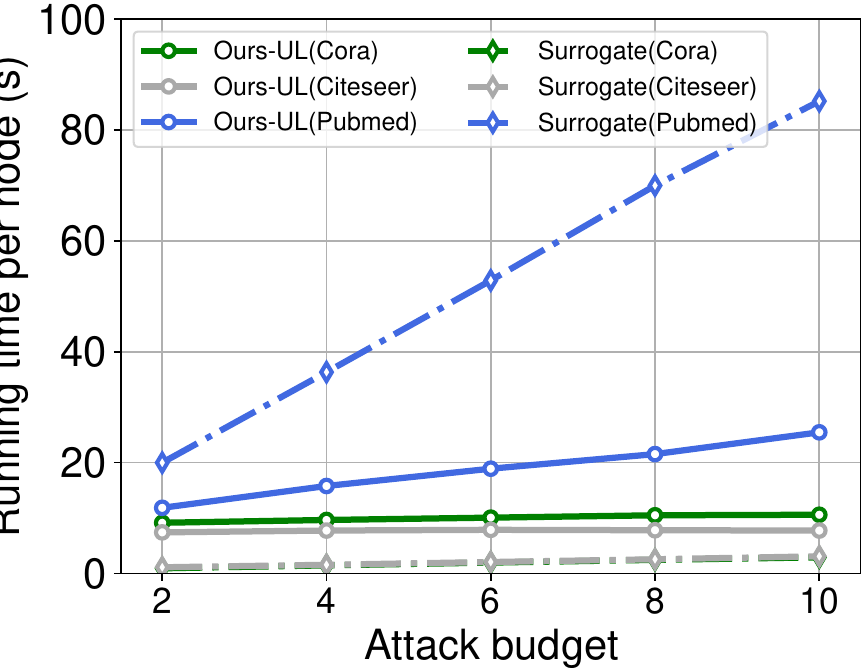}\label{fig:SGC_4layer_time}}
\vspace{-4mm}
\caption{(a) Attack success rate and (b) Running time of our attacks against four-layer SGC on the three graphs.}
%%\vspace{-5mm}
\end{figure}

\begin{table}[h]\renewcommand{\arraystretch}{1.0}
\centering
\caption{Transferability of our attacks against two-layer SGC to other GNNs on the three graphs. Attack budget is 6.}
%\vspace{-4mm}
% \small
\addtolength{\tabcolsep}{+2pt}
\begin{tabular}{|c|c|c|c|c|c|}
\hline
{\bf Dataset}   & {\bf Source}   & \multicolumn{4}{c|}{\bf Target} \\ \hline

\multirow{4}{*}{\bf Cora} & {\bf SGC} &  {\bf SGC}   &  {\bf GCN}   &    {\bf GAT} & {\bf JK-Net}    \\ \cline{2-6} 
&  {\bf No attack}   &  0   &  0.01   &  0.01   & 0.02     \\ \cline{2-6} 
&  {\bf Ours-KL}  & 0.99  & 0.78  & 0.73  & 0.71     \\ \cline{2-6}
&  {\bf Ours-UL}    & 0.99    & 0.75    & 0.72  & 0.74 \\ \hline \hline

\multirow{4}{*}{\bf Citeseer} & {\bf SGC} &  {\bf SGC}   &  {\bf GCN}   &    {\bf GAT} & {\bf JK-Net}    \\ \cline{2-6}  
&  {\bf No attack}   &  0   &  0.03   &  0.03   & 0.04     \\ \cline{2-6}
&  {\bf Ours-KL}  & 0.99  & 0.81    & 0.79  & 0.69     \\ \cline{2-6}    
&  {\bf Ours-UL} & 0.99    & 0.84    & 0.77  & 0.70  \\ \hline \hline

\multirow{4}{*}{\bf Pubmed} & {\bf SGC} &  {\bf SGC}   &  {\bf GCN}   &    {\bf GAT} & {\bf JK-Net}    \\ \cline{2-6} 
&  {\bf No attack}  &  0   &  0.04   &  0.07   & 0.07     \\ \cline{2-6}
&  {\bf Ours-KL}    &  0.89  & 0.83   &  0.76 & 0.81    \\ \cline{2-6}  
&  {\bf Ours-UL}    & 0.92   & 0.83    & 0.75  & 0.76 \\ \hline 
\end{tabular}
\label{tbl:transfer_2layer_SGC}
\end{table}

\begin{table}[!t]\renewcommand{\arraystretch}{1.0}
\centering
\caption{Comparing our attack vs. IG-FGSM~\cite{wu2019adversarial} on Cora.}
\addtolength{\tabcolsep}{+2pt}
\begin{tabular}{|c|c|c|c|c|c|c|c|}
\hline
{\bf Model} & {\bf budget}                 &  \bf 2     &   \bf  4   &  \bf 6     & \bf 8      &   \bf 10    &  \bf  Time    \\ \hline
\multirow{3}{*}{\bf SGC} & {\bf IG-FGSM} & 0.32 & 0.74 & 0.89 & 0.95 & 0.96 & 55s  \\ \cline{2-8}
   &  {\bf Nettack}             &   0.65    &   0.72    &  0.73     &  0.73    &  0.82     &   2.5s   \\ \cline{2-8}
   &  {\bf Ours}             &   0.90    &   0.93    &  0.95     &  0.95    &  0.96     &   0.1s   \\ \hline
\end{tabular}
\label{tbl:compared_with_igfgsm_sgc}
\end{table}

\begin{table}[t]\renewcommand{\arraystretch}{1.0}
\centering
\caption{
Comparing our attack vs. black-box attack \cite{wang2022bandits}. 
}
\addtolength{\tabcolsep}{+2pt}
\centering
\begin{tabular}{|c|c|c|c|c||c|c|c|c|c|} \hline 

{\bf SGC} & {\bf budget}  & {\bf 2} & {\bf 4} & {\bf 6}  & {\bf SGC}
 & {\bf budget}  & {\bf 2} & {\bf 4} & {\bf 6} \\ \hline
\multirow{2}{*}{\bf Cora} & {\bf Ours}  &  {0.81} & {0.92} & {0.98} & \multirow{2}{*}{\bf Citeseer} & {\bf Ours}  &  {0.90} & {0.96} & {0.98}  \\ \cline{2-5} \cline{7-10}
& {\bf \cite{wang2022bandits}}  &  {0.64} & {0.77} & {0.81}  & 
& {\bf \cite{wang2022bandits}}  &  {0.80} & {0.88} & {0.90} \\ \hline
\end{tabular}
\label{tbl:res_BBattack}
\end{table}

\end{document}